\pdfoutput=1
\documentclass[12pt]{article}
\usepackage[utf8]{inputenc}
\usepackage[T1]{fontenc}
\usepackage{amsmath}
\usepackage{amsfonts}
\usepackage{amssymb}
\usepackage{graphicx}
\usepackage{fullpage}
\usepackage{hyperref}
\usepackage{comment}
\usepackage{subcaption}
\usepackage[nosort]{cite}
\usepackage{bm}

\usepackage{tikz}
\usetikzlibrary{arrows}
\usetikzlibrary{shapes}
\usepackage{float}
\restylefloat{table}

\usepackage{MnSymbol}
\hypersetup{
	colorlinks=true,
	linkcolor=blue,
	filecolor=magenta,      
	urlcolor=blue,
}

\def\bP{{\bm P}}
\def\bX{{\bm X}}
\def\cN{{\mathcal N}}
\def\cO{{\mathcal O}}
\def\cD{{\mathcal D}}
\def\cI{{\mathcal I}}

\begin{document}

\title{Higher-Dimensional Symmetry of AdS$_2\times$S$^2$ Correlators}
\author{Theresa Abl, Paul Heslop and  Arthur E. Lipstein \vspace{7pt}\\ \normalsize \textit{
Department of Mathematical Sciences}\\\normalsize\textit{Durham University, Durham, DH1 3LE, United Kingdom}}
\maketitle
\begin{abstract} It was recently shown that IIB supergravity on AdS$_5\times$S$^5$ enjoys 10d conformal symmetry and that superstring theory on this background can be described using a 10d scalar effective field theory. In this paper we adapt these two complementary approaches to correlators of hypermultiplets in AdS$_2\times$S$^2$. In particular, we show that 4-point correlators of $1/2$-BPS operators in the 1d boundary can be computed using 4d conformal symmetry and a 4d effective action in the bulk. The 4d conformal symmetry is realised by acting with Casimirs of $SU(1,1|2)$, and is generically broken by higher derivative corrections. We point out similar structure underlying $\alpha'$ corrections to IIB supergravity in AdS$_5 \times$S$^5$. In particular, while the $\alpha'^3$ corrections can be written in terms of a sixth order Casimir acting on a 10d conformal block, similar structure does not appear in higher-order corrections. We note however that a specific combination of higher derivative corrections can give rise to Witten diagrams with higher dimensional symmetry at the integrand level, with breaking then arising from the measure.

\noindent 

\end{abstract}
\thispagestyle{empty}
\pagebreak
\tableofcontents

\section{Introduction}

Understanding quantum field theory and quantum gravity in curved backgrounds such as black holes and cosmological spacetimes is one of the most pressing questions in theoretical physics. The AdS/CFT correspondence is a very powerful tool for this purpose since it relates observables in curved backgrounds to correlation functions of a CFT in the boundary \cite{Maldacena:1997re}. Moreover, AdS is a useful toy model since AdS$_4$ is related by Wick rotation to dS$_4$ which is relevant for cosmology \cite{Maldacena:2002vr}, and AdS$_2\times$S$^2$ describes the near-horizon geometry of extremal black holes in four dimensions \cite{Bertotti:1959pf,Robinson:1959ev}. While a great deal of technology has been developed to compute CFT correlators, we would also like to be able to compute things directly in the bulk in order to have a complete grasp of the physics. Carrying out calculations directly in the bulk is very challenging however because of the complexity of Witten diagrams and worldsheet calculations. Nevertheless, much progress has been made for IIB string theory in AdS$_5 \times$ S$^5$. For example, it was discovered that supergravity in this background enjoys a 10d conformal symmetry, which is realised using 4-point correlators of protected operators in the boundary CFT, $\mathcal{N}=4$ SYM \cite{Caron-Huot:2018kta}\footnote{All 4-point correlators of protected operators in $\mathcal{N}=4$ SYM in the supergravity approximation were first computed in \cite{Rastelli:2017udc}.}. This property is inherited from scale-invariance of the corresponding flat space scattering amplitude after a conformal transformation to AdS$_5 \times$ S$^5$ and has very non-trivial consequences for $\mathcal{N}=4$ SYM correlators. Similar results were also found in AdS$_3 \times$ S$^3$ \cite{Rastelli:2019gtj,Giusto:2020neo,Aprile:2021mvq,Wen:2021lio} and AdS$_5 \times$ S$^3$ \cite{Alday:2021odx}. Subsequent work showed that 10d conformal symmetry is broken by string corrections \cite{Drummond:2020dwr}. In another development, a simple 10d scalar effective action was conjectured to describe all 4-point correlators of $1/2$-BPS operators in $\mathcal{N}=4$ SYM dual to tree-level string theory in AdS$_5 \times$ S$^5$ to all orders in $\alpha'$ \cite{Abl:2020dbx}. This action successfully reproduced previous results obtained using different methods~\cite{Goncalves:2014ffa,Alday:2018pdi,Drummond:2019odu,Drummond:2020dwr,Aprile:2020luw} and makes an infinite number of new predictions. 

In this paper we will develop these methods in the context of AdS$_2 \times$S$^2$. One motivation for doing so is to expand the toolbox for studying black holes in the real world. Another is that the correlators in this background have very simple mathematical structure, which can provide a toy model for better understanding 10d conformal symmetry in  AdS$_5 \times$S$^5$ and its breaking. On the other hand, much less is known about the AdS$_2$/CFT$_1$ correspondence due to various subtleties such as the absence of a stress tensor in a conventional 1d CFT \cite{Qiao:2017xif} and a large backreaction when matter is introduced in the bulk \cite{Maldacena:1998uz,Almheiri:2014cka,Maldacena:2016upp}. We will therefore consider a non-gravitational theory in the bulk. In particular, reducing $\mathcal{N}=8$ supergravity on AdS$_2\times$S$^2$ gives $\mathcal{N}=2$ graviton multiplets and hypermultiplets \cite{Michelson:1999kn,Lee:1999yu} and we will restrict our attention to 4-point tree-level correlators of the latter. We also study correlators of higher-derivative corrections to the interactions of hypermultiplets in AdS$_2 \times$S$^2$ analogous to $\alpha'$ corrections in AdS$_5 \times$S$^5$. From the point of view of the effective action, the two-derivative interactions that arise from reducing supergravity on AdS$_2 \times$S$^2$ correspond to a $\phi^4$ interaction and we refer to them as the 2-derivative sector. Higher-derivative interactions are then described by applying derivatives to $\phi^4$. In addition there is a generalised free theory described by non-interacting Witten diagrams in the bulk.

We can formally define correlators on the boundary which have $SU(1,1|2)$ superconformal and crossing symmetry. Using these symmetries, we can then deduce the structure of the lowest-charge correlators in the free, 2-derivative, and higher-derivative sectors. We then use 4d conformal symmetry to obtain all higher-charge correlators in the free and 2-derivative sectors, which correspond to Kaluza-Klein modes on the S$^2$, and match them with the predictions of the effective action. A key ingredient in the realisation of 4d conformal symmetry is the action of Casimirs of the superconformal group. On the other hand, we find that only a special infinite class of correlators derived from the effective action with higher derivative corrections can be reproduced by acting with Casimirs on 4d conformal blocks, suggesting that 4d conformal symmetry is generically broken by higher derivative corrections. This is further confirmed by the structure of anomalous dimensions of double trace operators which we compute using a procedure for unmixing operators which are degenerate at leading order in the central charge expansion reminiscent of~\cite{Aprile:2017xsp}. Whereas the anomalous dimensions are rational numbers in the 2-derivative sector and correspond to eigenvalues of a Casimir operator, they are generically irrational in higher derivative sectors.  

We find similar structure in 4-point correlators corresponding to $\alpha'$ corrections to supergravity in AdS$_5 \times$S$^5$. In particular, $\alpha'^3$ corrections are analogous to the 2-derivative sector in AdS$_2 \times$S$^2$ in that they are both described by a $\phi^4$ effective field theory and enjoy higher-d conformal symmetry. Similarly, $\alpha'^5$ corrections are analogous to 6-derivative corrections in AdS$_2 \times$S$^2$ and are shown to generically break 10d conformal symmetry, in the sense that the predictions of the 10d effective action cannot be matched by acting with Casimirs on 10d conformal blocks. We do  find a 6-derivative effective action for which the integrands of the Witten diagrams exhibit 10d conformal symmetry, but the resulting correlators  are in disagreement with $\cN=4$ SYM, specifically the coefficients fixed by  localisation in~\cite{Binder:2019jwn}. This provides a new point of view on the 10d conformal symmetry breaking.

This paper is organised as follows. In section \ref{1dcftreview} we review the formalism for 1d CFT correlators and the AdS$_2 \times$S$^2$ effective action and in section \ref{sec:freetheory} we compute free theory correlators and demonstrate their 4d conformal symmetry using Casimirs. In section \ref{sec:1dsupergravity}, we compute correlators in the 2-derivative sector and show agreement between the predictions of 4d conformal symmetry and the effective action. We also provide an intuitive explanation for the origin of this symmetry. Next, in section \ref{higherderv1} we compute correlators in higher derivative sectors and show that the 4d conformal symmetry is generically incompatible with the predictions of the effective action, except for a special infinite class of correlators. In section \ref{anomdim} we use an unmixing procedure to compute the anomalous dimensions of double trace operators in the 2- and 6-derivative sectors, finding further evidence for the breaking of 4d conformal symmetry by higher-derivative corrections. In section \ref{n=4}, we apply the lessons we have learned in AdS$_2 \times$S$^2$ to AdS$_5 \times$S$^5$, realising 10d conformal symmetry for $\alpha'^3$ corrections using Casimirs and providing a new perspective on the breaking of this symmetry by $\alpha'^5$ corrections. Finally, in section \ref{conclusion} we present our conclusions. We also have a number of Appendices providing more details about superconformal Casimirs, correlators of descendent operators, and further examples of unmixing.

\section{Overview}\label{1dcftreview}

In this section, we will describe some aspects of superconformal correlators and $SU(1,1|2)$ blocks in 1d. After that, we will review the effective action in AdS$\times$S and how to compute generalised Witten diagrams from this action.  

\subsection {1d Correlators} 
Note that it is not possible to define a stress tensor in 1d since any conserved current would simply be a constant. As a result, we only consider non-gravitational theories in the bulk whose correlators have superconformal symmetry arising from the isometries of the background, and can be formally thought of as correlators of a boundary CFT. It is also possible to recast the expansion of bulk correlators in terms of Newton's constant $G_N$ as an expansion of boundary correlators in terms of a formally defined central charge $c$. More precisely, we will associate $G_N \sim l_P^2$ with $1/c$, where $l_P$ is the Planck length. Free theory correlators are then $\mathcal{O}(c^0)$, while correlators in the 2-derivative sector correspond to $\mathcal{O}(1/c)$. In addition to the lack of a stress tensor in the boundary, another subtlety is that AdS$_2$ has two disconnected timelike boundaries. In this paper, we only consider correlators on a single boundary, but in principle one could consider correlators between the two boundaries \cite{Azeyanagi:2007bj}. While the former have singularities when points in the boundary collide, such singularities are absent in the latter. 

In 1d, 4-point conformal correlators can be expressed in terms of conformal cross ratios which take the following form:
\begin{equation}
z\bar{z}=u=\left(\frac{x_{12}x_{34}}{x_{13}x_{24}}\right)^{2},\,\,\,(1-z)(1-\bar{z})=v=\left(\frac{x_{23}x_{14}}{x_{13}x_{24}}\right)^{2}
\end{equation}
where $x_{ij}=x_{i}-x_{j}$. Solving for $z,\bar{z}$ in terms of
$u,v$ shows that 
\begin{equation}
z=\bar{z}=\frac{1}{2}(1+u-v)
\end{equation}
which follows from the fact that the discriminant $(1+u-v)^{2}-4u$
vanishes. Hence, conformal cross ratios in 1d correspond to a holomorphic limit of the standard cross ratios in higher dimensions. It is convenient to define $z=\bar{z}=x$ in terms of which cross
ratios are given by
\begin{equation}\label{1dand4dcrossratios}
u=x^{2},\qquad v=(1-x)^{2}.
\end{equation}

To describe superconformal correlators, we follow the formalism of \cite{Doobary:2015gia} (with $m=n=1$ in the notation of that paper) arising from analytic superspace~\cite{Galperin:1984av,Howe:1995md} and consider the super Grassmannian $Gr(1|1,2|2)$ of $(1|1)$ planes in $(2|2)$ dimensions. Coordinates on this Grassmannian can be given as
\begin{align}\label{coords}
X_i=\left(	\begin{array}{cc}
	x_i& \theta_i\\
	\bar \theta_i & y_i
	\end{array}
	\right)\ 
\end{align}
where $x_i$ is the (1d) space-time coordinate, $y_i$ is a (complex) internal coordinate used to deal with the $SU(2)$ structure and  $\theta_i,\bar \theta_i $ are Grassmann odd coordinates. Since we will be dealing with correlators of four operators on this space, we added a subscript $i=1,2,3,4$ to denote the particle number.

The 4d $\mathcal{N}=2$ hypermultiplet is the simplest to understand from a higher dimensional perspective. This multiplet on an AdS${}_2 \times$S${}^2$ background is dual to an infinite tower of fermionic 1d $1/2$-BPS multiplets~\cite{Michelson:1999kn,Lee:1999yu}. 
These are fermionic superfields of scaling dimension and $SU(2)$ rep $\Delta$. 
They can be written on the above super Grassmannian and decompose into the following component fields
\begin{align}\label{Psi}
	{\Psi}_{\Delta}(x_i, \theta_i,\bar \theta_i, y_i)
	=\psi_{\Delta}+
	 \theta_i \phi_{\Delta+\frac12}+ 
	  \bar \theta_i \bar \phi_{\Delta+\frac12}+  \theta_i \bar \theta_i \lambda_{\Delta+1}\qquad \Delta=\tfrac12,\tfrac32,\tfrac52,\dots
\end{align}
where in each case the subscript denotes the dimension of the field. The field $\psi_\Delta$ has $SU(2)$ rep $\Delta$, $\phi_\Delta$ has $SU(2)$ rep $\Delta-1$, $\lambda_\Delta$ has $SU(2)$ rep $\Delta-2$.  For the special multiplet $\Psi_{1/2}$ the descendant $\lambda_{3/2}$ is absent (it would have negative $SU(2)$ weight). 
Expanding in the $y_i$ coordinates manifests the $SU(2)$ indices for these reps
\begin{align}\label{comps}
	\psi_{\Delta}(x_i,y_i) &= \psi_{I_1\dots I_{2\Delta}}(x_i) y_i^{I_1}\dots y_i^{I_{2\Delta}}\qquad& \Delta&=\tfrac12,\,\tfrac32,\,\dots\notag\\
	\phi_{\Delta}(x_i,y_i) &= \phi_{I_1\dots I_{2\Delta-2}}(x_i) y_i^{I_1}\dots y_i^{I_{2\Delta-2}}\qquad& \Delta&=1,2,\dots\notag\\
	\lambda_{\Delta}(x_i,y_i) &= \lambda_{I_1\dots I_{2\Delta-4}}(x_i) y_i^{I_1}\dots y_i^{I_{2\Delta-4}}\qquad &\Delta&=\{\},\tfrac52,...\notag\\	
\end{align}
where $y_i^I=(1,y_i)$.

We will be interested in the 4-point functions of four $1/2$-BPS superconformal primaries:
\begin{equation}
\label{halfBPSfourpointprimanddesc}
	\langle  \psi_{\Delta_1}(x_1,y_1) \psi_{\Delta_2}(x_2,y_2)  \psi_{\Delta_3}(x_3,y_3) \psi_{\Delta_4}(x_4,y_4)\rangle=\hat P_{\Delta_i} G_{\psi_{\Delta_i}}(x,y),
\end{equation}
where
\begin{align}\label{prefactor}
\hat P_{\Delta_i}&=
g_{12}^{\Delta_1+\Delta_2} g_{34}^{{\Delta_3+\Delta_4}} 
\left(\frac{g_{24}}{g_{14}}\right)^{{\Delta_{21}}}
\left(\frac{g_{14}}{g_{13}}\right)^{{\Delta_{43}}}\\
g_{ij}&=\frac{y_{ij}}{x_{ij}}\ . \label{gijprop}
\end{align}
It will also be useful shortly to define the 2-point function of $\psi_{1/2}$ operators
\begin{equation}
\tilde{g}_{ij}=\frac{y_{ij}}{\left|x_{ij}\right|},
\label{gijdefferm}
\end{equation}
which is antisymmetric under $i \leftrightarrow j$, reflecting the fermionic statistics. The prefactor in \eqref{prefactor} by itself  transforms correctly as a $\langle  \psi_{\Delta_1}\psi_{\Delta_2}\psi_{\Delta_3}\psi_{\Delta_4}\rangle$ correlator under the bosonic subgroup, leaving a remaining function $G_\psi(x,y)$ which is invariant, and thus a function of 
$x,y$, the  space-time and internal cross-ratios:
\begin{align}\label{1dcrossratios}
x=\frac{x_{12}x_{34}}{x_{13}x_{24}}\qquad 	y=\frac{y_{12}y_{34}}{y_{13}y_{24}}\ .
\end{align}

Throughout this paper we will consider 4-point functions of $1/2$-BPS operators and in some cases also the correlators of descendants. The $1/2$-BPS operators are fermionic primaries $\psi_\Delta$ with dimensions and $SU(2)$ charges $\Delta=\tfrac12,\tfrac32,\dots$ However, it will turn out to be most useful to label the correlators in terms of the bosonic descendant operators $\phi_{\Delta+1/2}$ with dimensions $1,2,\dots$ Therefore, we will define the $1/2$-BPS operators as 
\begin{equation}\label{1dhalfbpsnorm}
\cO_p:=\left(-1\right)^{\frac{p}{2}}\frac{\psi_{p-\frac{1}{2}}}{\sqrt{2p-1}},
\end{equation}
where $p=1,2,\dots$ In this convention the half-BPS operators have dimension and $SU(2)$ representation $p-\frac{1}{2}$. Furthermore, it is useful to introduce the normalisation $\left(-1\right)^{p/2}(2p-1)^{-1/2}$ inspired by the higher-dimensional conformal symmetry~\cite{Caron-Huot:2018kta}. It will become clear why this is useful in the discussions below~\eqref{htpppp} and~\eqref{Lcorrelatorpqpq}. We then also label the descendants similarly, thus defining 
\begin{equation}
	L_p:=\left(-1\right)^{\frac{p}{2}}\frac{\phi_p}{\sqrt{2p-1}},
\end{equation}
so the superfield~\eqref{Psi} becomes 
\begin{align}\label{Ldescendant}
{\bf{O}}_p(x,y)=\cO_p(x,y)+\theta L_p(x,y)+\bar{\theta}\bar{L}_p(x,y)+..\ ,
\end{align}
where $L_p$ has dimension $p$ and $SU(2)$ charge $p-1$.

The correlator of primaries \eqref{halfBPSfourpointprimanddesc} is written:
\begin{align}\label{factoroutP}
	&\langle  \cO_{p_1}(x_1,y_1) \cO_{p_2}(x_2,y_2)   \cO_{p_3}(x_3,y_3) \cO_{p_4}(x_4,y_4)
	\rangle  =  P_{p_i} \times
	G_{p_1 p_2 p_3 p_4}(x,y)\ ,
	\end{align}
where the prefactor~\eqref{prefactor} becomes
\begin{align}
	P_{p_i}=\hat P_{(2p_i-1)/2}=
	g_{12}^{p_1+p_2-1} g_{34}^{p_3+p_4-1} 
	\left(\frac{g_{24}}{g_{14}}\right)^{{p_{21}}}
	\left(\frac{g_{14}}{g_{13}}\right)^{{p_{43}}}\ .
\end{align}
In this form, the superconformal Ward identities are simply  $\partial_x G(x,y)|_{x=y} = 0 $, i.e. $G(x,x)$ is independent of $x$ \cite{Doobary:2015gia}. This implies that we can write 
\begin{align}\label{kandHcorrelator}
&G_{p_i}(x,y) = k_{p_i}(x,y) + 	\frac{x-y}{x y} H_{p_i}(x,y)\ ,\notag\\
&\text{where}\  \ k_{p_i}(x,y)=\kappa \left(\frac{x}{y}\right)^{p_{43}}\ ,
\end{align}
with $H(x,x)$ finite  and with $k_{p_i}(x,y)$ defined such that it is annihilated by the super Casimir (defined later in~\eqref{Delta2def}),  $\mathcal{C}^{SU(1,1|2)}_{1,2}\left(P_{p_{i}}k_{p_{i}}\right)=0$ with a constant $\kappa$.
 
\subsection{Conformal Blocks}\label{sec:1dblocks}

In this subsection we discuss the conformal blocks in 1d. 4-point functions can be expanded in superconformal blocks $B_{\Delta,p}(x,y)$ as follows: 
  \begin{align}\label{1dblockexpansion}
&\langle  \cO_{p_1}(x_1,y_1) \cO_{p_2}(x_2,y_2)   \cO_{p_3}(x_3,y_3) \cO_{p_4}(x_4,y_4)
\rangle  \nonumber \\&= 
\sum_{\Delta=1}^\infty\sum_{p=0}^{p_1+p_2-2}
A^{p_1p_2p_3p_4}_{\Delta, p}  \,
g_{12}^{p_1+p_2-1} g_{34}^{p_3+p_4-1} 
\left(\frac{g_{24}}{g_{14}}\right)^{{p_{21}}}
\left(\frac{g_{14}}{g_{13}}\right)^{{p_{43}}}
\, B_{\Delta,p,p_{12},p_{34}}(x,y)\ ,
\end{align}
where $p_{ij}=p_i-p_j$, $g_{ij}$ is defined in \eqref{gijprop}, and the coefficient $A_{\Delta,p}$ is given in terms of a sum of squares of OPE coefficients
\begin{align}
&A^{p_1p_2p_3p_4}_{\Delta,p}= \sum_{\cO^{\Delta,p},\tilde \cO^{\Delta,p}}
C_{p_1p_2}^{\cO}   C_{p_3p_4}^{\tilde \cO}  C_{\cO \tilde \cO}\ .
\end{align}

The superconformal blocks were derived for a general conformal group $SU(m,m|2n)$ in~\cite{Doobary:2015gia}. Specialising to $m=n=1$, the blocks for long multiplets are given by
\begin{align}\label{1dlongblocks}
	 B^\text{long}_{\Delta,p,p_{12},p_{34}}(x,y) =&-(x-y)(-x)^{\Delta} {}_2F_1(\Delta+1-p_{12},\Delta+1+p_{34};2\Delta+2;x)\notag\\
	 &\times y^{-p-1} {}_2F_1(-p+p_{12},-p-p_{34};-2p;y)\ .
\end{align}
For half-BPS multiplets $\Delta=p$, and the blocks are
\begin{align}\label{1dshortblocks}
B^\text{half-BPS}_{p,p_{12},p_{34}}(x,y)=&\left(-\frac{x}{y}\right)^p \Big(1+(x-y) \sum_{i=1}^k [x^{-i} \, _2F_1(p+1-i-p_{12},p+1-i+p_{34};2p+2-2 i;x) ]\nonumber\\  
&\times y^{i-1}\, _2F_1(i-p+p_{12},i-p-p_{34};2 i-2 p;y)\Big)\ ,
\end{align}
where $k=\min(p-p_{12},p+p_{34})$ and the square bracket means we take the regular piece as $x\rightarrow0$.

It is useful to note that the block of an operator with dimension $\Delta$ and $SU(2)$ representation $p$ contributes as follows to the 4-point function
	  \begin{align}\label{cpwapprox}
	&\langle  \cO_{p_1}(x_1,y_1) \cO_{p_2}(x_2,y_2)   \cO_{p_3}(x_3,y_3) \cO_{p_4}(x_4,y_4)
	\rangle  \nonumber \\&\sim
	\sum_{\Delta,p}
	A^{p_1p_2p_3p_4}_{\Delta,p}  \,
	g_{12}^{p_1+p_2-1} g_{34}^{p_3+p_4-1} 
	\left(\frac{g_{24}}{g_{14}}\right)^{{p_{21}}}
	\left(\frac{g_{14}}{g_{13}}\right)^{{p_{43}}}
		(-x)^{\Delta}y^{-p}\left(1+\cO(x,y)\right)	\ ,
	\end{align}
 where the higher orders in $x,y$ correspond to spacetime or $SU(2)$ descendants.

\subsection{4d Effective Action}\label{sec:4dscalarSeff}

In~\cite{Abl:2020dbx} a 10d scalar effective action was introduced which generates all half-BPS 4-point correlators in $\cN=4$ SYM described by tree-level string theory in AdS$_5\times$S$^5$. In AdS$\times$S the effective action takes the form
\begin{align}\label{4dSeff}
	V^{\text{AdS$\times$S}}(\phi)=
	\frac{G_N}{4!}\Big[&A\phi^4+l_s^4\Big(3 B(\nabla\phi.\nabla\phi)^2+ 6 C \nabla^{2}\nabla_{\mu}\phi\nabla^{\mu}\phi\phi^{2}\Big) \notag\\
&+l_s^6\Big(6 D (\nabla\phi.\nabla\phi)(\nabla_\mu\nabla_\nu\phi\nabla^\mu\nabla^\nu\phi)+6E \nabla_{\mu}\nabla^{2}\nabla_{\nu}\phi\nabla^{\mu}\nabla^{\nu}\phi\phi^{2}\Big)+\dots\Big]\, 
\end{align}
where $G_N$ is Newton's constant and $l_s$ is the analogue of the string length. These parameters can be written in terms of the AdS radius $R$ as follows: 
\begin{equation}
G_{N}=R^{2}/c,\,\,\,l_{s}^{4}=aR^{4},
\end{equation}
where $R$ is the AdS radius, and $c$ and $a$ are analogous to the central charge and string coupling, respectively. Some of the terms in the effective action can be uplifted directly from the flat space action. It is important to note however that this uplift is not unique: the action contains ambiguities because the covariant derivatives no longer commute with each other leading to curvature corrections which vanish in the flat space limit. Furthermore the coefficients $B,C,..$ themselves can have an expansion in the dimensionless parameter $l_P/R^2$:
\begin{align}\label{4dcoeffexpansion}
	&B(a) =B_0+B_1\tfrac{l_s^2}{R^2}+\dots\notag \\
	&C(a) =C_0+C_1\tfrac{l_s^2}{R^2}+\dots\notag \\
	&\dots
\end{align}
For simplicity, we will set $R=1$ from now on throughout this paper, but it will be understood that terms suppressed by $1/R$ vanish in the flat space limit. Whereas in AdS$_5 \times$S$^5$ case the $\phi^4$ interaction describes $\alpha'^3$ corrections to supergravity, in AdS$_2\times$S$^2$ it describes the 2-derivative sector obtained from dimensional reduction of supergravity, as we will see in section \ref{sec:sugra4dconfsym}. Hence, the effective action in AdS$_2\times$S$^2$ treats the 2-derivative sector on equal footing with higher-derivative corrections. 

\subsection{AdS$\times$S Witten Diagrams}

Using the effective action, we can compute generalised Witten diagrams that treat AdS and the sphere on equal footing. This is explained in detail in \cite{Abl:2020dbx}, so we will just state the main formulae. The generalised bulk-to-boundary propagator in AdS$_{d+1} \times$S$^{d+1}$ is
\begin{align}\label{bulktoboundary10d}
	G(\hat X,\hat Y ; X,Y)=\left(-2\hat{X}. X-2\hat{Y}. Y\right)^{-\Delta}\ ,
\end{align}
where $\left\{ \hat{X},\hat{Y}\right\} $ are embedding coordinates for AdS and sphere, respectively, written in terms of $(d+2)$-dimensional Minkowski and Euclidean space, respectively, and which satisfy $\hat{X}^2=-\hat{Y}^{2}=-1$. Similarly, $\left\{ X,Y\right\} $ are boundary coordinates satisfying $X^{2}=Y^{2}=0$. These are related to $d$-dimensional boundary coordinates via $-2X_{i}\cdot X_{j}=x_{ij}^{2}$ and $-2Y_{i}\cdot Y_{j}=y_{ij}^{2}$, where $\left\{ i,j\right\} $ label a pair of points. Note that the sphere does not actually have a boundary since it is compact, but the $Y$ coordinates naturally describe $1/2$-BPS operators in the boundary of AdS. When $\Delta=d$, the generalised bulk-to-boundary propagator satisfies the equations of motion for a massless scalar field:
\begin{equation}
\nabla^2 G=0.
\end{equation}
Since the effective action in AdS$\times S$ describes a massless 4d scalar field, we will therefore set $\Delta=d$ in practice. Note that the bulk-to-boundary propagators have a normalisation
\begin{equation} \label{cdelta}
 \mathcal{C}_\Delta=\frac{\Gamma\left(\Delta\right)}{2\pi^{d/2}\Gamma\left(\Delta-d/2+1\right)}. 
\end{equation}
We will dress the generalised Witten diagrams with this normalisation to obtain correlators.  

A generalised contact Witten diagram is then defined as
\begin{align}\label{adsxcont}
	D^{\text{AdS}_{d+1}\times \text{S}^{d+1}}_{\Delta_1\Delta_2\Delta_3\Delta_4}(X_i,Y_i) = \frac1{(-2)^{2\Sigma_{\Delta}}}\int_{\text{AdS}\times\text{S}}  \frac{d^{d+1} \hat X d^{d+1} \hat Y}{(P_1+Q_1)^{\Delta_1}(P_2+Q_2)^{\Delta_2}(P_3+Q_3)^{\Delta_3}(P_4+Q_4)^{\Delta_4}}\ ,
\end{align}
where we introduce the shorthand
\begin{align}
	P_i=\hat X.X_i\ , \qquad Q_i=\hat Y.Y_i\ .
\end{align}
It is now straightforward to expand this AdS$\times$S contact diagram into an infinite number of standard AdS contact diagrams multiplied by spherical analogues. In particular, using
\begin{align}\label{propexp}
	\frac1{(P+Q)^{\Delta}}=\sum_{p=0}^\infty(-1)^p \frac{(p+1)_{\Delta-1}}{\Gamma(\Delta)}\frac{Q^p}{P^{p+\Delta}}
\end{align}
four times gives the expansion:%
\begin{align}\label{DinDB}
	D^{\text{AdS}\times\text{S}}_{\Delta_1\Delta_2\Delta_3\Delta_4}(X_i,Y_i) 
	=& \sum_{p_i=0}^\infty 
	\prod_{i=1}^4(-1)^{p_i}\frac{(p_i+1)_{\Delta_i-1}}{\Gamma(\Delta_i)} \notag\\
	&\times
	D^{(d)}_{p_1+\Delta_1,p_2+\Delta_2,p_3+\Delta_3,p_4+\Delta_4}(X_i) 
        B^{(d)}_{p_1p_2p_3p_4}(Y_i) \ ,
\end{align}
where
\begin{align}\label{adscont}
	D^{(d)}_{\Delta_1\Delta_2\Delta_3\Delta_4}(X_i) = \frac1{(-2)^{2\Sigma_{\Delta}}}\int_\text{AdS}  \frac{d^{d+1} \hat X}{(\hat X.X_1)^{\Delta_1}(\hat X.X_2)^{\Delta_2}(\hat X.X_3)^{\Delta_3}(\hat X.X_4)^{\Delta_4}}\ ,
\end{align} 
\begin{align}\label{scont}
	B^{(d)}_{p_1p_2p_3p_4}(Y_1,Y_2,Y_3,Y_4)= (-2)^{2\Sigma_p}\int_{\text{S}} {d^{d+1} \hat Y}   (\hat Y.Y_1)^{p_1}(\hat Y.Y_2)^{p_2}(\hat Y.Y_3)^{p_3}(\hat Y.Y_4)^{p_4} \ ,
\end{align}
with $\Sigma_\Delta=(\Delta_1{+}\Delta_2{+}\Delta_3{+}\Delta_4)/2$ and $\Sigma_p=(p_1{+}p_2{+}p_3{+}p_4)/2$.
For more details, see section 2.3 of \cite{Abl:2020dbx}. It is convenient to define normalised functions
\begin{align}\label{normalisedDandB}
	D^{(d)}_{\Delta_1\Delta_2\Delta_3\Delta_4}(X_i)= \cN^{\text{AdS}_{d+1}}_{\Delta_i} D_{\Delta_1\Delta_2\Delta_3\Delta_4}(X_i)  \ ,\,\,\,  B^{(d)}_{p_1p_2p_3p_4}(Y_i) =   \cN^{\text{S}^{d+1}}_{p_i} B_{p_1p_2p_3p_4}(Y_i),
\end{align}
where
\begin{align}
    \cN^{\text{AdS}_{d+1}}_{\Delta_i}=\frac{\tfrac12 \pi^{d/2} \Gamma(\Sigma_\Delta-d/2)}{(-2)^{\Sigma_\Delta}\prod_i\Gamma(\Delta_i)}\ ,\,\, \cN^{\text{S}^{d+1}}_{p_i}={2\times 2^{\Sigma_p}}\frac{\pi^{d/2+1}\prod_i\Gamma(p_i{+}1)}{\Gamma(\Sigma_p{+}d/2{+}1)} \ .
\end{align}
It is also conventional to write Witten diagrams in terms of $\bar{D}$ functions which are closely related to the D-functions described above. For more details, see for example Appendix D of \cite{Arutyunov:2002fh}.

Note that the contact diagrams in \eqref{adsxcont} come from bulk interaction vertices without derivatives. More generally, contact diagrams with derivatives take the form
\begin{align}\label{decoration}
 \frac1{4!}\frac{1}{(-2)^{2\Sigma_\Delta}}\int_{\text{AdS}\times\text{S}} {d^{d+1} \hat X} {d^{d+1} \hat Y} \times\left( \prod_i \frac{Q_i^{n^Q_i}P_i^{n^P_i}\times (\Delta_i)_{n_i}}{(P_i+Q_i)^{\Delta_i+n_i}}\right)\times \left(\prod_{i<j}(X_i.X_j)^{n^X_{ij}}(Y_i.Y_j)^{n^Y_{ij}}\right)  \ ,
\end{align}
with $n_i = n_i^P+n_i^Q+\sum_j n^X_{ij}+\sum_j n^Y_{ij}$. Expanding them as above then gives 
\begin{align}\label{MinDB}
(-2)^{2\Sigma_X+2\Sigma_Y}	\sum_{p_i=0}^\infty \left(\prod_{i=1}^4(-1)^{p_i}\frac{(p_i+1)_{\Delta_i+n_i-1}}{\Gamma(\Delta_i)} D^{(d)}_{p_i+\Delta_i+n_i-n^P_i}(X_i)B_{p_i+n^Q_i}(Y_i)\right)\left(\prod_{i<j}(X_i.X_j)^{n^X_{ij}}(Y_i.Y_j)^{n^Y_{ij}}\right)\ ,
\end{align}
where $\Sigma_X,\Sigma_Y$ represent the sum of all the $n^X_{ij},n^Y_{ij}$, respectively.

\section{Free theory}\label{sec:freetheory}

In this section we will derive 1d free theory correlators and show that they are consistent with 4d conformal symmetry. The free theory correlators can be computed from Wick contractions. For equal charge operators we obtain
\begin{align}
&\langle\cO_{p}(x_{1},y_{1})\cO_{p}(x_{2},y_{2})\cO_{p}(x_{3},y_{3})\cO_{p}(x_{4},y_{4})\rangle|_{c^0}\notag\\
&=\frac{1}{\cN^{(0)}_{pppp}}\left[(\tilde{g}_{12}\tilde{g}_{34})^{2p-1}-(\tilde{g}_{13}\tilde{g}_{24})^{2p-1}+(\tilde{g}_{14}\tilde{g}_{23})^{2p-1} \right]\notag
\\
&= \frac{1}{\cN^{(0)}_{pppp}}\left(\tilde{g}_{12}\tilde{g}_{34}\right)^{2p-1}\left[1+\left(\frac{x}{y}\right)^{2p-1}\left(\left(-{\rm sgn}x\right)+\left(\frac{1-y}{1-x}\right)^{2p-1}{\rm sgn}\left[x\left(1-x\right)\right]\right)\right]\ , \label{eq39}
\end{align}
where $\tilde{g}$ is defined in \eqref{gijdefferm},
\begin{align}
\cN^{(0)}_{p_i}=\left(-1\right)^{\Sigma_p}\sqrt{(2p_1-1)(2p_2-1) (2p_3-1) (2p_4-1)}
\end{align}
comes from the normalisation in~\eqref{1dhalfbpsnorm} and $\Sigma_p=\tfrac{1}{2}\left(p_1+p_2+p_3+p_4\right)$.
Choosing our external points such that $x_1<x_2<x_3<x_4$, then $\tilde g_{12}=g_{12}, \tilde g_{34}=g_{34}$ and   $ 0 \le x \le 1$ and so comparing to \eqref{factoroutP}, we find that
\begin{align}\label{Gfreepppp}
G^{(0)}_{pppp}(x,y)  =\frac1{\cN^{(0)}_{pppp}} \left[ 1 + \left(\frac xy \right)^{2p-1}\left(-1+ \left(\frac {1-y}{1-x} \right)^{2p-1} \right)\right]\, 
\end{align}
where the superscript indicates free theory correlators. According to \eqref{kandHcorrelator} this decomposes into
\begin{align}\label{kHpppp0}
k^{(0)}_{pppp}(x,y)=\frac1{\cN^{(0)}_{pppp}} \,, \,\ H^{(0)}_{pppp}(x,y) =\frac1{\cN^{(0)}_{pppp} } \frac{x\,y}{x-y}\left(\frac{x}{y}\right)^{2p-1}\left[-1+\left(\frac{1-y}{1-x}\right)^{2p-1}\right]\ .
\end{align}

Similarly, for unequal charges $\{p_i\}=pqpq$, where $p>q$ we obtain
\begin{align}
&\langle \cO_p(x_1,y_1)\cO_q(x_2,y_2)\cO_p(x_3,y_3)\cO_q(x_4,y_4)\rangle_{c^0}=-\,\frac1{\cN^{(0)}_{pqpq} }\,\tilde{g}^{2p-1} _{13}\tilde{g}^{2q-1} _{24}\ ,  \\ 
&\text{and}\ G^{(0)}_{pqpq}(x,y)=-\,\frac1{\cN^{(0)}_{pqpq} }\left(\frac{x}{y}\right)^{p+q-1}\ , \label{Gfreepqpq}
\end{align}
where the operators have dimensions and charges $(2p-1)/2,\,(2q-1)/2$ and we decompose $G^{(0)}_{pqpq}$ into 
\begin{equation}\label{Hkfreepqpq}
k^{(0)}_{pqpq}(x,y)=-\frac1{\cN^{(0)}_{pqpq} }\left(\frac{x}{y}\right)^{q-p}\ ,\  \  H^{(0)}_{pqpq}(x,y)=\frac1{\cN^{(0)}_{pqpq} }\frac{x\,y}{x-y} \left(\frac{x}{y}\right)^{q-p}\left[1-\left(\frac{x}{y}\right)^{2p-1}\right]\ .
\end{equation}

\subsection{4d Conformal Symmetry}\label{sec:freetheory4dconfsym}

To realise 4d conformal symmetry in the free theory, we need to consider the correlators of descendants $L_p,\bar L_p$ rather than the super primaries $\cO_p$. Remarkably, as we show in appendix~\ref{sec:desc},  the two correlators are related by a simple second order differential operator which is closely related to the quadratic Casimir of the superconformal group acting on points 1 and 2:
\begin{align}
\label{descendentcorrelator1}
\langle L_{p_1}L_{p_2}\bar L_{p_3} \bar L_{p_4} \rangle={\mathcal I}^{-1}\mathcal{C}^{SU(1,1|2)}_{1,2} \langle \cO_{p_1}\cO_{p_2}\cO_{p_3}\cO_{p_4} \rangle\ ,
\end{align}
where 
\begin{align}\label{I}
	{\mathcal I} = x_{12}x_{34}y_{13}y_{24}-y_{12}y_{34}x_{13}x_{24} = (x-y)x_{13}x_{24}y_{13}y_{24},
\end{align}
and
${\mathcal C}_{1,2}^{SU(1,1|2)}$ denotes the superconformal quadratic Casimir:
\begin{align}\label{Delta2def}
{\mathcal C}_{1,2}^{SU(1,1|2)}&= P_{p_i} \times \frac{x-y}{xy}   \Delta^{(2)}      \frac{xy}{x-y}    P_{p_i}^{-1}\ , \notag \\
\Delta^{(2)} &= {\mathcal D}^+_x - {\mathcal D}^-_y \ ,\notag\\
{\mathcal D}^\pm_z &=z^2 \partial_z(1-z)\partial_z\pm(p_{12}+p_{43})z^2\partial_z - p_{12}p_{43} z\ .
\end{align}
For a derivation of the quadratic Casimir, see Appendix \ref{casimirderivation}.

In the decomposition given by \eqref{kandHcorrelator} the Casimir only sees $H_{p_i}$: 
\begin{align}\label{casonfreetheory}
\mathcal{C}^{SU(1,1|2)}_{1,2} \langle \cO_{p_1}\cO_{p_2}\cO_{p_3}\cO_{p_4} \rangle = P_{p_i} \frac{x-y}{xy}  \Delta^{(2)} H_{p_i}(x,y)= P_{p_i} \frac{x-y}{xy}  \tilde H_{p_i}(x,y)\ ,
\end{align}
where we define
\begin{align}
\label{Htdef}
\tilde H_{p_i}(x,y) =\Delta^{(2)} H_{p_i}(x,y)\ .
\end{align}
The correlator in \eqref{descendentcorrelator1} is then given by
\begin{align}\label{Lcorrelator}
	  \langle L_{p_1}L_{p_2}\bar L_{p_3} \bar L_{p_4} \rangle= \frac {P_{p_i}}{x_{12}x_{34} y_{12}y_{34}} \tilde H_{p_i}(x,y)\ .
	 \end{align}

Let us first consider the case of equal charges given in \eqref{kHpppp0}. The action of $\Delta^{(2)}$ then yields 
\begin{align}\label{htpppp}
\tilde H_{pppp} = \frac {x^{2p}}{y^{2p-2}}\left(1+\frac{(1-y)^{2p-2}}{(1-x)^{2p}}\right)\ .
\end{align}  
Note that acting with $\Delta^{(2)}$ on $H$ gives a factor of $(2p-1)^2$ which gets cancelled by the normalisation $1/\cN^{(0)}_{pppp}$. This is how the normalisation in~\eqref{1dhalfbpsnorm} was chosen, up to the sign which will become clear shortly. From~\eqref{Lcorrelator} and~\eqref{htpppp} we get
\begin{equation}\label{Lcorrelatorpppp}
\langle L_p L_p \bar L_p\bar L_p\rangle=g_{14}^{2p-2}g_{23}^{2p-2}\frac{1}{x^2_{14} x^2_{23}}+g_{13}^{2p-2}g_{24}^{2p-2}\frac{1}{x_{13}^2 x_{24}^2}\ .
\end{equation} 
Note that the correlator of bosonic descendants is given in terms of bosonic 2-point functions $g_{ij}$~\eqref{gijprop}. For the lowest case $p=1$ we have
 \begin{align}
 	\langle L_1 L_1\bar L_1  \bar L_1 \rangle = \frac1{x_{14}^2x_{23}^2}+\frac{1}{x_{13}^2x_{24}^2}  \ .
 \end{align}
To see the 4d conformal symmetry, now lift this to four dimensions by replacing $x_{ij}^2 \rightarrow x_{ij}^2+y_{ij}^2= x_{ij}^2(1+g_{ij}^2)$ which gives
 \begin{align}\label{LLLbLb4d}
 \langle LL\bar L \bar L \rangle_{c^0}  &=  \left[\frac1{x_{14}^2x_{23}^2}+\frac{1}{x_{13}^2x_{24}^2}\right]_{4d} \notag \\
 &=\frac1{x_{14}^2x_{23}^2} \frac 1 {(1+g_{14}^2)(1+g_{23}^2)}+\frac{1}{x_{13}^2x_{24}^2} \frac 1 {(1+g_{13}^2)(1+g_{24}^2)}\ ,
 \end{align}
where we have defined the generator of all operators $L_p$ 
\begin{align}\label{Lgen}
	L=\sum_{p=1}^\infty L_p\ .
\end{align}
This correlator is thus a 4d object which contains all 1d free theory correlators. In particular, if we expand $	(1+g^2)^{-1}=1-g^2 +g^4-g^6+...$ and keep the two terms proportional to $g_{14}^{2p-2} g_{23}^{2p-2}$ and $g_{13}^{2p-2} g_{24}^{2p-2}$, this indeed reproduces the prediction in~\eqref{Lcorrelatorpppp}.
 
As another example, let us consider correlators with unequal charges in \eqref{Hkfreepqpq}. Acting on this with $\Delta^{(2)}$ gives
\begin{equation}
\tilde{H}_{pqpq}(x,y)=\left(-1\right)^{p+q}\frac{x^{p+q}}{y^{p+q-2}}\ .
\end{equation}
Using~\eqref{Lcorrelator} then yields
 \begin{equation}\label{Lcorrelatorpqpq}
\langle L_p L_q \bar L_p\bar L_q\rangle=\left(-1\right)^{p+q} g_{13}^{2p-2}g_{24}^{2q-2}\frac{1}{x^2_{13} x^2_{24}}\ ,
\end{equation}
 which agrees with the term proportional to $g_{13}^{2p-2} g_{24}^{2q-2}$ in the expansion of the generating function~\eqref{LLLbLb4d}. Thus the choice of signs in our normalisation of the half-BPS operators~\eqref{1dhalfbpsnorm} was inspired by the realisation of the 4d conformal symmetry in free theory, similarly to the rest of the normalisation.

\section{2-derivative sector}\label{sec:1dsupergravity}

In this section, we will derive all 4-point correlators corresponding to two-derivative interactions of bulk hypermultiplets. There is an infinite tower of such correlators corresponding to Kaluza-Klein modes on the $S^2$. The basic strategy will be to first deduce the lowest charge correlator from superconformal symmetry and crossing, and then uplift it to arbitrary charges using the assumption of 4d conformal symmetry. This result will then be confirmed using Witten diagrams derived from the bulk effective action. Finally, we will show that the 4d uplift of the lowest charge correlator corresponds to a 4d conformal block, and provide a physical interpretation in terms of the partial wave expansion of the 4d bulk scattering amplitude arising in the flat space limit. This will have implications for anomalous dimensions computed in section \ref{anomdim}.

\subsection{Lowest-Charge Correlator} \label{sec:sugra1111int}

In this subsection we determine the correlator for $p_i=1$ using crossing symmetry and various other constraints. Note that crossing symmetry in 1d is subtle. For previous treatments see \cite{Mazac:2018qmi,Giombi:2017cqn}. We make the following ansatz in terms of bosonic 2-point functions $g_{ij}$~\eqref{gijprop}\footnote{Although we are using bosonic 2-point functions, the 4-point function is still a valid solution to the Ward identities with the correct crossing  symmetry properties.}:
\begin{align}\label{ansatzsugra1111}
\langle\cO_{1}(x_{1},y_{1})\cO_{1}(x_{2},y_{2})\cO_{1}(x_{3},y_{3})\cO_{1}(x_{4},y_{4})\rangle|_{1/c} =g_{12}g_{34}\frac{x}{y}(x-y)a(x)\ .
\end{align}
If we exchange positions $1$ and $3$, which takes $x \rightarrow 1-x$ and $y \rightarrow 1-y$, and set the result equal to minus the original expression (by Fermi statistics) we find the crossing constraint
\begin{equation}\label{sugrabosoniccrossing1}
a(1-x)=a(x)\ .
\end{equation}
 Let us now consider exchanging $2$ with $3$, which takes $x \rightarrow 1/x$ and $y \rightarrow 1/y$:
\begin{equation}\label{ansatz23}
\langle\cO_{1}(x_{1},y_{1})\cO_{1}(x_{3},y_{3})\cO_{1}(x_{2},y_{2})\cO_{1}(x_{4},y_{4})\rangle|_{1/c}=g_{12}g_{34}\frac{x}{y}(x-y)\left(-\frac{1}{x^2}a(1/x)\right)\ .
\end{equation} 
Equating this with minus the original expression then implies
\begin{equation} \label{sugrabosoniccrossing2}
a\left(\frac1{x}\right)=x^2 a(x)\ .
\end{equation}
Finally, for the third crossing condition, consider exchanging $1$ with $3$ in~\eqref{ansatz23}:
\begin{align}
\langle\cO_{1}(x_{3},y_{3})\cO_{1}(x_{1},y_{1})\cO_{1}(x_{2},y_{2})\cO_{1}(x_{4},y_{4})\rangle|_{1/c}\notag=g_{12}g_{34}\frac{x}{y}(x-y)\frac{1}{(1-x)^2}a\left(\frac1{1-x}\right)\ .
\end{align}
Equating this with the expression in~\eqref{ansatzsugra1111} then gives the condition
\begin{equation}
a\left(\frac1{1-x}\right)=(1-x)^2 a(x) \ ,
\end{equation}
which follows from~\eqref{sugrabosoniccrossing1} and~\eqref{sugrabosoniccrossing2}. 

Finally we consider the asymptotic structure of the correlator as $x \rightarrow 0$ arising from knowledge of the OPE. Since the operators are fermionic there is no exchange of the identity operator and the lowest weight operator appearing has $\Delta=1$. More precisely, consider the conformal block expansion in \eqref{1dblockexpansion} and \eqref{cpwapprox}
 \begin{align}
&\langle  \cO_{1}(x_1,y_1) \cO_{1}(x_2,y_2)   \cO_{1}(x_3,y_3) \cO_{1}(x_4,y_4)
\rangle  \nonumber \\
&= \sum_{\Delta=1}^\infty A^{1111}_{\Delta, 0}  \, g_{12} g_{34}\, B^\text{long}_{\Delta,0}(x,y)\nonumber \\
&\sim \sum_{\Delta=1}^\infty
A^{1111}_{\Delta, 0}  \,
g_{12} g_{34}\, (-x)^\Delta \left(1+\cO(x)\right),
\end{align}
and compare to~\eqref{ansatzsugra1111} in the $x\rightarrow 0$ limit. This implies that  $a(x)$  must satisfy the additional constraint
\begin{equation} \label{constraint3}
a(x)=\mathcal{O}(1)\,
\end{equation}
as $x \rightarrow 0$ (neglecting terms containing $\log(x)$ arising from the perturbative expansion).

We now make the following ansatz for  $a(x)$ based on the perturbative OPE structure
\begin{align}\label{ansatza}
a(x)= p(x) \log x^2 + p(1-x){\log (1-x)^2} +r(x) \ ,
\end{align}
where
\begin{equation} \label{ansatz1}
p(x)=\frac{1}{x-1}\frac{1}{x^{k}(1-x)^{k}}\sum_{i=0}^{m}b_{i}x^{i},\,\,\,r(x)=\frac{1}{x-1}\frac{1}{x^{k}(1-x)^{k}}\sum_{i=0}^{m}c_{i}x^{i},\,\,\,r(x)=r(1-x),
\end{equation}
and $k$ and $m$ are integers. This ansatz is natural since the holomorphic limit of $\bar{D}$ functions takes this form. We then find the minimal solution to the constraints in ~\eqref{sugrabosoniccrossing1},~\eqref{sugrabosoniccrossing2} and~\eqref{constraint3}:
\begin{equation}\label{2dervsol}
a(x)= \frac{\log x^2}{1-x}+\frac{\log (1-x)^2}{x}=-\bar{D}^\text{hol}_{1111}(x)\ ,
\end{equation}
where  $\bar{D}^\text{hol}$ is the holomorphic limit of the $\bar{D}$ function.  We will obtain all higher-charge correlators in the 2-derivative sector from 4d conformal symmetry in the next subsection.

\subsection{4d Conformal Symmetry}\label{sec:sugra4dconfsym}

Decomposing the lowest-charge correlator in \eqref{ansatzsugra1111} and \eqref{2dervsol} according to~\eqref{kandHcorrelator} gives 
\begin{equation}\label{1dsugracorrelator}
H_{1111}^{(2)}=-u \bar{D}^\text{hol}_{1111},\qquad  k_{1111}^{(2)}=0,
\end{equation}
where the superscript ${(2)}$ indicates that we are considering 2-derivative interactions in the bulk. We can obtain all higher charge correlators in this sector using 4d conformal symmetry. First lift the 1d cross-ratios and holomorphic $\bar{D}$-function to the usual cross-ratios and $\bar{D}$-functions, using the relations~\eqref{1dand4dcrossratios} for the cross-ratios. To get an object that transforms as a 4d conformal correlator, divide by ${x_{12}^2x_{34}^2}$ and view all Lorentz invariants as living in 4d. We then obtain
\begin{align}\label{4d2deriv}
	\langle \cO \cO \cO \cO\rangle_{1/c}= \cI \times \left[\frac{H^{(2)}_{1111}}{x_{12}^2x_{34}^2}\right]_{4d},\ 
\end{align}
where $\cO$ is the generator of all half BPS operators 
\begin{align}\label{ogen}
	\cO= \sum_{p=1}^\infty \cO_p\ .
\end{align}
By the subscript $4d$ we mean that all  Lorentz invariants live in 4d, so replace  $x_{ij}^2\rightarrow x_{ij}^2(1+g_{ij}^2)$:
\begin{equation}\label{sugra4dlift}
\left[\frac{H^{(2)}_{1111}}{x_{12}^2x_{34}^2}\right]_{4d}=\left[\frac{-u\bar{D}_{1111}(u,v)}{x_{12}^2x_{34}^2}\right]_{4d}\ =\  \frac{-u_{4d}\bar{D}_{1111}(u_{4d},v_{4d})}{x_{12}^2x_{34}^2}\frac{1}{(1+g_{12}^2)(1+g_{34}^2)}\ ,
\end{equation}
where
\begin{align}
u_{4d}=\frac{x_{12}^2 x_{34}^2}{x_{13}^2 x_{24}^2}\frac{(1+g_{12}^2)(1+g_{34}^2)}{(1+g_{13}^2)(1+g_{24}^2)}\ ,\ \ v_{4d}=\frac{x_{14}^2 x_{23}^2}{x_{13}^2 x_{24}^2}\frac{(1+g_{14}^2)(1+g_{23}^2)}{(1+g_{13}^2)(1+g_{24}^2)}\ .
\end{align}

This 4d object in \eqref{4d2deriv} then encodes all higher-charge correlators. To obtain correlators with specific charges, expand in small $g_{ij}$ and take the coefficients of the appropriate powers of $g_{ij}^2$. Before deriving the  general formula for any correlator, let us consider the explicit expansion for a few examples. For the first few cases with charges $\{p_i\}=ppqq$ this looks like:
\begin{alignat}{3}
\langle \cO_1 \cO_1 \cO_1 \cO_1 \rangle_{1/c} =&\cI \times g_{12} g_{34}\times &&\left(- \bar{D}_{1111}\right)\ , \nonumber \\
\langle \cO_1 \cO_1 \cO_2 \cO_2 \rangle_{1/c} =&\cI \times g_{12} g_{34}\times && g_{34}^2\left(u \bar{D}_{1122}\right)\ , \nonumber \\
\langle \cO_1 \cO_1 \cO_3 \cO_3  \rangle_{1/c} =&\cI \times g_{12} g_{34}\times &&\tfrac{1}{2}g_{34}^4\left(-u\, \bar{D}_{1133}\right)\ , \nonumber \\
\langle \cO_2 \cO_2 \cO_2 \cO_2  \rangle_{1/c} =&\cI \times g_{12} g_{34}\times &&\left(u\, g_{12}^2 g_{34}^2 +g_{13}^2 g_{24}^2 +v\, g_{14}^2 g_{23}^2 \right)\left(-u\, \bar{D}_{2222}\right)\ , \nonumber \\
\langle \cO_2 \cO_2 \cO_3 \cO_3  \rangle_{1/c} =&\cI \times g_{12}g_{34}\times &&\left(\tfrac{1}{2}u\, g_{12}^2 g_{34}^4+g_{13}^2g_{24}^2 g_{34}^2+v\, g_{14}^2 g_{23}^2 g_{34}^2\right)\left(u\,\bar{D}_{2233}\right)\ , \nonumber \\
\langle \cO_3 \cO_3 \cO_3 \cO_3  \rangle_{1/c} =&\cI \times g_{12}g_{34}\times &&\left(\tfrac{1}{4}u^2 g_{12}^4 g_{34}^4+\tfrac{1}{4} g_{13}^4 g_{24}^4+ \tfrac{1}{4} v^2 g_{14}^4 g_{23}^4+u\, g_{12}^2 g_{34}^2 g_{13}^2 g_{24}^2\right.\nonumber \\
& &&\left. +u\,v\, g_{12}^2 g_{34}^2 g_{14}^2 g_{23}^2+v\, g_{13}^2 g_{24}^2 g_{14}^2 g_{23}^2 \right)\left(-u\,\bar{D}_{3333}\right)\ .
\label{dbarexpansion}
\end{alignat}
Note that after Taylor expanding, the right hand side of these equations is written directly in terms of 1d kinematics and so the $\bar D$ functions are automatically taken in the holomorphic limit.
Using the decomposition in~\eqref{kandHcorrelator} we find: 
\begin{align}\label{ppqqupliftedsugra}
H^{(2)}_{1122}&=u\bar{D}^{\text{hol}}_{1122} \ ,\quad H^{(2)}_{1133}=- \tfrac{1}{2}u\bar{D}^{\text{hol}}_{1133} \ ,\quad H^{(2)}_{2222}=- 2 u^2\frac{1-y+y^2}{y^2}\bar{D}^\text{hol}_{2222} \ ,\nonumber \\
H^{(2)}_{2233}&= \tfrac{1}{2}u^2 \frac{4-4y+3y^2}{y^2}\bar{D}^\text{hol}_{2233} \ ,\quad H^{(2)}_{3333}=-\tfrac{3}{2}u^3 \frac{(1-y+y^2)^2}{y^4}\bar{D}^\text{hol}_{3333}\ .
\end{align}

There is a simple formula for all charges which is easiest to see if  one first writes the uplifted 4-point correlator \eqref{sugra4dlift} in terms of normalised D-functions in \eqref{normalisedDandB}:
\begin{equation}
\left[\frac{H^{(2)}_{1111}}{x_{12}^2x_{34}^2}\right]_{4d}=\left[\frac{-u\bar{D}_{1111}}{x_{12}^2x_{34}^2}\right]_{4d}=\left[-D_{1111}\right]_{4d}\ .
\end{equation}
The 4d $D$-function is a function of 4d $[x_{ij}^2]_{4d} = x_{ij}^2(1+g_{ij}^2)$ and thus, for example, the coefficient of $\left(g_{12}^2\right)^n$ is
\begin{equation}
\frac{\left(x_{12}^2\right)^n}{n!}\frac{d^n}{d \left(x_{12}^2\right)^n}D_{\Delta_1\Delta_2\Delta_3\Delta_4}=\frac{\left(-x_{12}^2\right)^n}{n!}D_{\Delta_1+n\, \Delta_2+n\, \Delta_3 \Delta_4}
\end{equation}
where we used the relation \cite{Arutyunov:2002fh} 
\begin{equation}
\frac{d}{d\,x_{12}^2}D_{\Delta_1\Delta_2\Delta_3\Delta_4}=-D_{\Delta_1+1\, \Delta_2+1\, \Delta_3\Delta_4}\ .
\end{equation}
Hence a general correlator with charges $p_i$ is given by
\begin{align}\label{4dliftsugraallcorr}
\left[\frac{H^{(2)}_{p_1 p_2 p_3 p_4}}{x_{12}^2 x_{34}^2}\right]_{4d}&=\sum_{\{d_{ij}\}}\prod_{i<j}\left(\left(-1\right)^{d_{ij}}\left(g_{ij}^2\right)^{d_{ij}}\frac{\left(x_{ij}^2\right)^{d_{ij}}}{d_{ij}!}\right) (-D_{p_1 p_2 p_3 p_4})\notag\\
&=\sum_{\{d_{ij}\}}\prod_{i<j}\left(\frac{\left(y_{ij}^2\right)^{d_{ij}}}{d_{ij}!}\right) \left(-1\right)^{\Sigma_p+1}D_{p_1 p_2 p_3 p_4}\notag\\
&=(-1)^{\Sigma_p+1} D_{p_1 p_2 p_3 p_4}\times B_{p_1-1\,p_2-1\,p_3-1\,p_4-1}\ ,
\end{align}
where the normalised B-function is defined in \eqref{normalisedDandB} and
\begin{align}
\sum_{i<j} d_{ij}=p_i-1\, ,\ 0\leq d_{ij}=d_{ji}\, ,\  d_{ii}=0\ .
\end{align}
Hence, expanding the 4d formula~\eqref{4d2deriv}, we find that correlators with arbitrary charge in the 2-derivative sector are given by
\begin{align}
		\langle \cO_{p_1}\cO_{p_2}\cO_{p_3}\cO_{p_4} \rangle_{1/c} = \cI \times (-1)^{\Sigma_p+1} D_{p_1 p_2 p_3 p_4}\times B_{p_1-1\,p_2-1\,p_3-1\,p_4-1}\ .
\end{align}
We will derive this formula from the 4d effective action in the next subsection.

Finally it will be  useful to understand the implications of the 4d symmetry for the correlator of descendants $\langle LL \bar L  \bar L\rangle$.
Notice that~\eqref{4d2deriv} with \eqref{descendentcorrelator1} implies that 
\begin{align}
	\langle LL\bar L\bar L\rangle_{1/c} = \cI^{-1}	\mathcal{C}^{SU(1,1|2)}_{1,2} \langle \cO \cO \cO \cO\rangle_{1/c}=  \cI^{-1}	\mathcal{C}^{SU(1,1|2)}_{1,2} \cI \times \left[\frac{H^{(2)}_{1111}}{x_{12}^2x_{34}^2}\right]_{4d}\ .
\end{align}
Now note that $\cI^{-1}	\mathcal{C}^{SU(1,1|2)}_{1,2} \cI$ is simply equal to the Casimir for the maximal bosonic subgroup $SO(1,2)\times SO(3)$ of $SU(1,1|2)$ (see appendix~\ref{casimirderivation} and in particular~\eqref{casimirs}). 
So we see that the 4d symmetry manifests itself for the correlator $\langle LL\bar L\bar L\rangle_{1/c}$ as
\begin{align}\label{lllblb2}
	\langle LL\bar L\bar L\rangle_{1/c} =\mathcal{C}^{SO(1,2)\times SO(3)}_{1,2}  \left[\frac{H^{(2)}_{1111}}{x_{12}^2x_{34}^2}\right]_{4d}\ .
\end{align}
In other words the generator of the correlator of descendants is a $SO(1,2)\times SO(3)$ Casimir acting on a 4d object. 

\subsection{Effective Action}\label{sec:sugraeffectiveaction}

All 4-point half-BPS correlators in the 2-derivative sector can be computed from the first term in the effective action~\eqref{4dSeff}, which is simply a $\phi^4$ interaction:
\begin{align}
 S_2 = \frac1{4!}\, A \times \int_{\text{AdS}\times\text{S}} {d^{2} {\hat X}}{d^{2} {\hat Y}} \phi(\hat X,\hat Y)^4\ .
\end{align}
Using the generalised bulk-to-boundary propagators in~\eqref{bulktoboundary10d} we obtain the  
AdS$_2\times$S$^2$ Witten diagram for this $\phi^4$ contact interaction, leading to the following correlators (where recall $\cO$ generates all the $1/2$-BPS operators~\eqref{ogen}):
\begin{align}\label{sugrawittendiagram}
\langle\cO\cO\cO\cO \rangle^{(2)}_\text{int} &=  \frac1{4!}\,A\,	 \frac{(\mathcal{C}_1)^4}{(-2)^{4}}\int_{\text{AdS}\times\text{S}}  
\frac{{d^{2} {\hat X}}{d^{2} {\hat Y}}}{(P_1+Q_1)(P_2+Q_2)(P_3+Q_3)(P_4+Q_4)}\notag \\
&=\frac{1}{4!}\,A\,(\mathcal{C}_1)^4\times D_{1111}^{\text{AdS}_2\times\text{S}^2}\ .
\end{align}
Recall that $P_i=\hat{X}.X_i$, $Q_i=\hat{Y}.Y_i$, $\mathcal{C}_\Delta$ is defined in \eqref{cdelta}, and the AdS$\times$S $D$-functions are defined in~\eqref{adsxcont}.  

To extract correlators with specific charges one expands in the appropriate powers in $Y_i$ using the Taylor expansion of the 4d bulk-to-boundary propagators
\begin{align}\label{expand4dprop}
	(P_i+Q_i)^{-1} = \sum_{p=1}^\infty (-1)^{p-1}(P_i)^{-p} (Q_i)^{p-1}\ .
\end{align}
We then obtain%
\footnote{This is~\eqref{DinDB} with $\Delta_i=d=1$ and with $p_i\rightarrow p_i-1$ to account for the fact that the lowest correlator is labelled with $p_i=1$ rather than $p_i=0$. We do not need to worry about the minus signs in the factors $(-1)^p$ in~\eqref{expand4dprop} since $B_{p_1 p_2 p_3 p_4}=0$ if $p_1+p_2+p_3+p_4$ is odd.}:
\begin{align}\label{sugraallcorrSeff}
 &\langle \cO_{p_1} \cO_{p_2} \cO_{p_3} \cO_{p_4} \rangle_\text{int}^{(2)}\notag\\ &= \frac1{4!}\,A\,\frac{(\mathcal{C}_1)^4}{(-2)^{4}} \int_{\text{AdS}_2} {d^2\hat X} \prod_i\frac1{(P_i)^{p_i}} \times\int_{\text{S}^2} {d^2\hat Y} \prod_i(Q_i)^{p_i-1}\notag \\
&= A'\,\left(-1\right)^{\Sigma_p+1}   D_{p_1 p_2 p_3 p_4} \times B_{p_1-1\,p_2-1\,p_3-1\,p_4-1} \ ,
\end{align} 
which matches the result we obtained using 4d conformal symmetry in \eqref{4dliftsugraallcorr}.

\subsection{Physical Interpretation} \label{interp} 

In this section we will clarify the origin of 4d conformal symmetry. So far we found that both the free theory~\eqref{LLLbLb4d} and 2-derivative sector~\eqref{4d2deriv} and~\eqref{lllblb2} have a four-dimensional origin. Both cases are in fact written in terms of 4d conformal 4-point functions for operators with dimension 1. Any such conformal 4-point function has a simple expansion in terms of 4d bosonic conformal blocks of the form~\cite{Dolan:2000ut}: 
\begin{equation}\label{4dblockexp}
	\langle LL\bar L \bar L \rangle = 	 \frac{1}{x_{12}^2x_{34}^2}
\sum_{\Delta,l}A_{\Delta,l}\,
G_{\Delta,l}(u,v)\ ,
\end{equation}
where 
\begin{align}\label{4dconformalblocks}
G_{\Delta,l}(z,\bar{z})=&\frac{1}{z-\bar{z}}u^{\frac12(\Delta-l)} \big(\left(-\tfrac12 z\right)^l z  \, _2F_1\left(\tfrac{1}{2} (\Delta+l),\tfrac{1}{2} (\Delta +l);\Delta +l;z\right)\notag \\
&\times\, _2F_1\left(\tfrac{1}{2}(\Delta-l-2),\tfrac{1}{2}(\Delta-l-2);\Delta-l-2;\bar{z}\right)-\left(z\leftrightarrow \bar{z}\right)\big)\  .
\end{align}

For the free theory correlator in~\eqref{LLLbLb4d}, the 4d conformal block expansion is then given by
\begin{equation}
u+\frac{u}{v}  = \sum_{n,l}A^{(0)}_{\Delta_0,l}\,
G_{\Delta_0,l}(u,v)\ 
\end{equation}
and we only find non-zero coefficients if  $\Delta_0=l+2$:
\begin{equation}
A^{(0)}_{l+2,l}=\frac{2^{1+l} (l!)^2}{(2\,l)!}.
\end{equation}
This can be understood by recalling that this is the block expansion of a 4d  scalar field of dimension 1, which is therefore massless. So the only operators one can construct out of its  OPE are the spin $l$ currents $L \partial^l L$.

Next let's do a 4d conformal block expansion of the $\log u$ part of the 4d correlator describing  the 2-derivative sector~\eqref{4d2deriv},\eqref{lllblb2}:
\begin{equation}
-u\bar{D}_{1111}|_{\log u}=\sum_{\Delta_{0},l}A_{\Delta_{0},l}^{(0)}\gamma_{\Delta_{0},l}\,G_{\Delta_{0},l}(u,v).
\end{equation}
We find that only the 4d block with spin-0 contributes,  with $\Delta_0=2$ and anomalous dimension $\gamma_{2,0}=1$. 
In other words the $\log u$ part of the 2-derivative sector is simply a single spin-0 block $G_{2,0}$ so (using~\eqref{4d2deriv})
\begin{align}\label{spin0block1d}
	\langle \cO \cO \cO \cO\rangle_{1/c}= \cI \times \left[\frac{G_{2,0}(u,v)\log u +..}{x_{12}^2x_{34}^2}\right]_{4d}\ .
\end{align}
This agrees with the  partial wave coefficient of the corresponding flat space scattering amplitude, as we will now explain.

Let us first review how this works in $\mathcal{N}=4$ SYM and then adapt it to the present case. In \cite{Caron-Huot:2018kta}, it was observed that in the flat space limit supergravity correlators in AdS$_5 \times$S$^5$ reduce to the following scattering amplitude:
\begin{equation}
A_{10}=\frac{G_{N}\delta^{16}(Q)}{stu}\rightarrow G_{N}\frac{s^{3}}{tu},
\label{10damplitude}
\end{equation}
where we have taken the dilaton component. The 10d conformal symmetry of the corresponding $\mathcal{N}=4$ SYM correlators is then inherited from the flat space scattering amplitude after noting that $G_{N}\delta^{16}(Q)$
is dimensionless and AdS$_5 \times$S$^5$ is conformally flat. It is then possible to relate the anomalous dimensions deduced from these correlators to partial wave coefficients of the scattering amplitude.

In more detail, the partial wave expansion is given
by
\begin{equation}
A_{d}(s,\cos\theta)=\frac{1}{s^{(d-4)/2}}\sum_{l}(l+1)_{d-4}C_{l}(\cos\theta)\mathcal{A}_{l}^{d}(s)\label{partialwaveexpansion}
\end{equation}
where $\cos\theta=1+\frac{2t}{s}$. Writing \eqref{10damplitude} in
terms of $s$ and $\theta$, we find
\begin{equation}
A_{10}(s,\cos\theta)=\frac{4G_{N}s}{\sin^{2}\theta}.
\end{equation}
The single power of $s$ in the numerator indicates 2-derivative interactions,
as expected for supergravity. Comparing this to \eqref{partialwaveexpansion}
implies that the partial wave coefficients are schematically 
\begin{align}\label{10dpartialwavecoeff}
\mathcal{A}_{l}^{10}(s)\sim1+\frac{R^8}{c}\frac{s^{4}}{\left(l+1\right)_{6}}\ ,
\end{align}
where Newton's constant $G_N\sim R^8/c$ in 10d, $R$ is the AdS radius, and $c$ is the central charge. Remarkably, this has the same form as the $\mathcal{N}=4$ SYM anomalous dimensions first derived in \cite{Aprile:2017xsp} after unmixing operators whose scaling dimensions are degenerate at leading order in the $1/c$ expansion:
\begin{align}\label{10deffectivespinintro}
e^{\frac1{c}\gamma^\text{(2)}_{4d}}\sim1+\frac1{c}\frac{\delta^{(8)}}{\left(l+1\right)_{6}},\,
\end{align}
where $\delta^{(8)}$ is a certain factor corresponding to the eigenvalues of an eighth order differential operator which corresponds to a Casimir of the superconformal group $SU(2,2|4)$. 

Let us now adapt this discussion to AdS$_2 \times$S$^2$. In this case, the flat space amplitude for hypermultiplets with 2-derivative interactions is given by \cite{Rastelli:2019gtj}
\begin{equation}
A_{4}=G_{N}\delta^{4}(Q)\rightarrow G_{N}s,
\end{equation}
where we have taken the scalar component. Note that $G_{N}\delta^{4}(Q)$
is dimensionless in 4d, so this amplitude has 4d conformal symmetry.
The factor of $s$ indicates 2-derivative interactions. In contrast to the 10d amplitude, it has no $\theta$ dependence which implies that only $l=0$ contributes to the partial wave expansion, in agreement with the conformal block analysis above. Hence, the partial wave coefficients are given by
\begin{equation}
\mathcal{A}_{l=0}^{4}(s)\sim 1+\frac{R^2}{c}\,s\ ,
\end{equation}
where $G_N\sim R^2/c$ in 4d. This argument suggests that the corresponding anomalous dimensions obtained after unmixing degenerate operators in the 1d CFT should be given by
\begin{equation}\label{predictiondelta2}
e^{\tfrac1{c}\gamma_{1d}}\sim1+\frac{1}{c}\,\delta^{(2)}\ ,
\end{equation}
where $\delta^{(2)}$ is a second-order superconformal Casimir. We will verify this in section \ref{anomdim}.

\section{Higher-derivative corrections} \label{higherderv1} 

In this section, we will generalise the analysis of the previous section to 4-point correlators describing higher-derivative interactions of hypermultiplets in AdS$_2\times$S$^2$. First we deduce the lowest-charge correlators from crossing and relate them to the holomorphic limit of analogous correlators in AdS$_{d>2}$ found in \cite{Heemskerk:2009pn}. Then we compute higher-charge correlators using the effective action and explore the extent to which they exhibit 4d conformal symmetry by trying to reproduce them by acting with superconformal Casimirs on 4d conformal blocks. We find that this is not generically possible except for a special class of correlators, suggesting that 4d conformal symmetry is broken by higher-derivative interactions in the bulk.

\subsection{Lowest-Charge Correlators}\label{sec:higherderivcrossing}

In this section we will generalise the $\left\langle \mathcal{O}_{1}\mathcal{O}_{1}\mathcal{O}_{1}\mathcal{O}_{1}\right\rangle  $ correlator in the two-derivative sector in \eqref{1dsugracorrelator} to higher-derivative interactions in the bulk. Following the method of section \ref{sec:sugra1111int}, we make the ansatz
\begin{equation}
\left\langle \mathcal{O}_{1}\mathcal{O}_{1}\mathcal{O}_{1}\mathcal{O}_{1}\right\rangle  =g_{12}g_{34}(x/y)(x-y)a(x)
\end{equation}
where
\begin{equation}
a(1-x)=a(x),\,\,\,a(1/x)=x^{2}a(x),\,\,\,a(x)=\mathcal{O}(1),
\label{constraintsahigherderv}
\end{equation}
as required by crossing symmetry and the constraint that the correlator doesn't encode the exchange of the identity operator (for more details, see section \ref{sec:sugra1111int}). Taking $a(x)$ of the form in \eqref{ansatza} and \eqref{ansatz1}, we then find that \eqref{constraintsahigherderv} only has solutions for $k=2q$ where $q$ is a non-negative integer, and $m=3k$. The $k=0$ solution is the one previously found in the 2-derivative sector in \eqref{1dsugracorrelator}. 

Further insight into these solutions can be obtained by computing the average anomalous dimensions of operators contributing to their conformal block expansion. The average anomalous dimensions are obtained from the conformal block expansion of the part of the correlator proportional to $\ln u$: 
\begin{equation}
\label{acpw}
\left.a(x)\right|_{\ln u}=\sum_{\Delta=1}A_{\Delta}^{(0)}\gamma_{\Delta}F_{\Delta}(x),\,\,\,F_{\Delta}(x)=x^{\Delta-1}\,_{2}F_{1}(\Delta+1,\Delta+1,2\Delta+2,x),
\end{equation}
where we expand in terms of standard non-supersymmetric 1d conformal blocks which follow from setting $p_i=1$ in \eqref{1dlongblocks}. The OPE coefficients are obtained by expanding the free theory correlator in \eqref{Gfreepppp} with $p=1$ in terms of these blocks and are given by
\begin{equation}
\label{freeope}
A_{\Delta}^{(0)}=2\left(\Delta!\right)^{2}/\left(2\Delta\right)!,\,\,\,\Delta\in{\rm odd.}
\end{equation}
Plugging \eqref{freeope} into \eqref{acpw}, we then find that the anomalous dimensions $\gamma_\Delta$ scale like $\Delta^{2k}=\Delta^{4q}$ as $\Delta \rightarrow \infty$ for the solutions found in the previous paragraph. In \cite{Heemskerk:2009pn} it was shown that the large-$\Delta$ scaling of anomalous dimensions is tied to the number of derivatives in the corresponding bulk interactions. Since the $k=0$ solution corresponds to 2-derivative bulk interactions, it follows that solutions with general $k$ correspond to interactions with $2k+2=4q+2$ derivatives. Furthermore, using the decomposition in \eqref{kandHcorrelator}, solutions corresponding to $(4q+2)$-derivative interactions in the bulk can be written as 
\begin{equation}\label{higherderivlowestcharge}
H^{(4q+2)}_{1111}=x^2 a(x) =-u\,(1+u^q+v^q)\bar{D}^\text{hol}_{q+1\,q+1\,q+1\,q+1},
\end{equation}
modulo lower-derivative solutions.

Let us explain the relation to higher-derivative 4-point interactions in AdS$_{d>2}$. In that case, the solutions to the crossing
equations for the dual CFT correlators are labelled by spin $l=0,2,4,...$, and for spin $l$ there
are $l/2+1$ solutions corresponding to interaction vertices of a bulk effective action with $2l,2l+2,...,3l$ derivatives \cite{Heemskerk:2009pn}. Reducing these solution to AdS$_{2}$ by taking the holomorphic limit, one finds that
the $l/2+1$ solutions collapse to a single solution, modulo lower-spin solutions \cite{Ferrero:2019luz}. This solution corresponds to a $2l$ derivative interaction i.e. lowest number of derivatives
in the spin-$l$ tower of solutions, and is consistent with the result obtained
in the previous paragraph. For example in AdS$_{d>2}$ the solutions for $l=0,2$ are given by
\begin{align}
\left\langle 1111\right\rangle _{l=0,\partial^{0}}=& \bar{D}_{1111},\notag\\
\left\langle 1111\right\rangle _{l=2,\partial^{4}}=&(1+u+v)\bar{D}_{2222},\notag\\
\left\langle 1111\right\rangle _{l=2,\partial^{6}}=&\bar{D}_{3232}+\bar{D}_{2323}+u^{2}\bar{D}_{3322}+u\bar{D}_{2233}+v^{2}\bar{D}_{2332}+v\bar{D}_{3223},
\end{align}
where the subscript indicates the spin and number of derivatives in the bulk effective action, and we have taken the scaling dimensions of the external operators to be $\Delta=1$.
Taking the holomorphic limit of the 6-derivative solution then gives
a linear combination of the 4-derivative and $l=0$ solutions:
\begin{equation}
\left.\left\langle 1111\right\rangle _{l=2,\partial^{6}}\right|_{hol}=\frac{17}{7}(1+u+v)\left.\bar{D}_{2222}\right|_{hol}-\frac{1}{7}\left.\bar{D}_{1111}\right|_{hol}.
\end{equation}
Hence, in the holomorphic limit there is only a 4-derivative solution at $l=2$ plus a zero derivative solution. In the present context, these correspond to 2-derivative and 6-derivative interactions of bulk hypermultiplets. 

\subsection{Effective Action}\label{sec:fourderiveffectiveaction}

As explained in section \ref{sec:sugraeffectiveaction}, two-derivative interactions of the bulk theory are described by a $\phi^4$ interaction in the effective action. Higher derivative corrections are then obtained by applying covariant derivatives to the $\phi^4$ vertex. 4-point correlators can then be computed from generalised Witten diagrams in terms of unfixed coefficients of the effective action. 

In the previous subsection, we showed that the lowest higher derivative corrections appear at six derivatives, which correspond to four-derivative corrections in the effective action. In this subsection we will use the effective action to compute the associated 1d CFT correlators for arbitrary charges of the external operators. As shown in \cite{Abl:2020dbx}, there are only two linearly independent terms in the 4-derivative effective action (corresponding to six derivative corrections to the Lagrangian for bulk hypermultiplets):
\begin{equation}\label{fourderivSeff}
S_6=B_0 S_6^\text{main}+C_0 S_6^\text{amb}\ ,
\end{equation}
with
\begin{align}\label{effact4deriv}
S_6^\text{main}&=\frac{3}{4!}\int_{\text{AdS}\times\text{S}}d^2\hat{X}d^2\hat{Y}\left(\nabla\phi.\nabla\phi\right)\left(\nabla\phi.\nabla\phi\right)\ ,\notag\\
S_6^\text{amb}&=\frac{6}{4!}\int_{\text{AdS}\times\text{S}}d^2\hat{X}d^2\hat{Y}\nabla^2\nabla_\mu\phi\nabla^\mu\phi\,\phi^2\ .
\end{align}
We refer to the second interaction as an ambiguity because it vanishes in the flat space limit. In this limit we can commute the $\nabla^2$ to the right, and then use $\nabla^2 \phi=0$. The Witten diagram for the main interaction is
\begin{align}\label{4derivmaincontactdiagram}
 &\langle \cO \cO \cO \cO\rangle^{(6)}_\text{main}\notag\\
 & =\frac{1}{4!}\frac{(\mathcal{C}_1)^4}{(-2)^4}\int_{\text{AdS}\times\text{S}}d^2\hat{X}d^2\hat{Y}\frac{N_{12}N_{34}+N_{13}N_{24}+N_{14}N_{23}}{\left(P_1+Q_1\right)^2 \left(P_2+Q_2\right)^2 \left(P_3+Q_3\right)^2 \left(P_4+Q_4\right)^2}\ ,
\end{align}
where
\begin{align}
N_{ij}&=X_i.X_j+Y_i.Y_j+P_iP_j-Q_iQ_j\ .
\end{align}
Furthermore, the Witten diagram for the ambiguity is
\begin{align}\label{4derivambcontactterm}
 &\langle \cO \cO \cO \cO\rangle ^{(6)}_\text{amb}\notag\\
 & =-\frac{1}{4!}\frac{(\mathcal{C}_1)^4}{(-2)^4}\int_{\text{AdS}\times\text{S}}\frac{d^2\hat{X}d^2\hat{Y}}{\prod_i\left(P_i+Q_i\right)}\sum_{i<j}\frac{L_{ij}}{\left(P_i+Q_i\right)\left(P_j+Q_j\right)}\ ,
\end{align}
where
\begin{align}
L_{ij}&=X_i.X_j-Y_i.Y_j+P_iP_j+Q_iQ_j\ .
\end{align}

Using \eqref{decoration} and \eqref{MinDB}, we can expand~\eqref{4derivmaincontactdiagram} and~\eqref{4derivambcontactterm} in terms of $D$- and $B$-functions, and then take the holomorphic limit of the  $\bar{D}$-functions. Using the decomposition in \eqref{kandHcorrelator}, the explicit expressions for some of the correlators are:
\begin{align}\label{fourdercorrelators}
H_{1111}^\text{(6)}=&\,3 u\left(\bar{D}^\text{hol}_{1111}-5\left(1+u+v\right)\bar{D}^\text{hol}_{2222}\right)\ ,\notag\\
H_{pp11}^\text{(6)}=&\frac{\left(-1\right)^{p+1}u^p}{(p-1)!} \left(f_1(p)\bar{D}^\text{hol}_{pp11}+f_2(p)\,u\,\bar{D}^\text{hol}_{p+1\,p+1\,11} +f_3(p)\left(1+u+v\right)\bar{D}^\text{hol}_{p+1\,p+1\,22}\right)\ ,\notag\\
H_{p1p1}^\text{(6)}=&\frac{\left(-1\right)^{p+1}u^{\frac{p+1}{2}}}{(p-1)!\,y^{p-1}} \left(f_1(p)\bar{D}^\text{hol}_{p1p1}+f_2(p)\,\bar{D}^\text{hol}_{p+1\,1\,p+1\,1} +f_3(p)\left(1+u+v\right)\bar{D}^\text{hol}_{p+1\,2\,p+1\,2}\right)\ ,\notag\\
H_{p11p}^\text{(6)}=&\frac{\left(-1\right)^{p+1}u^{\frac{p+1}{2}}}{(p-1)!\,y^{p-1}} \left(f_1(p)\bar{D}^\text{hol}_{p11p}+f_2(p)\,\bar{D}^\text{hol}_{p+1\,11\,p+1} +f_3(p)\left(1+u+v\right)\bar{D}^\text{hol}_{p+1\,22\,p+1}\right)\ ,\notag\\
H_{2222}^\text{(6)}=&\,2\frac{u^2}{y^2}\,\Big(\left[-8\,C_0\,\left(1-y+y^2\right)+2\left(41-6y+6y^2\right)\right]\bar{D}^\text{hol}_{2222}+35\,u\left(y^2-1\right)\bar{D}^\text{hol}_{3322} \notag\\
&+35\left(y-2\right)y\bar{D}^\text{hol}_{3223}-63\left(1+u+v\right)\left(1-y+y^2\right)\bar{D}^\text{hol}_{3333}\Big)\ ,
\end{align}
where
\begin{align}
f_1(p)&=2\,C_0\,p(p-1)+p^2\,(4\,p^2-8\,p+1)\ ,\ f_2(p)=-(1+2p)(2\,p^2-3\,p+1)\ ,\notag\\
f_3(p)&=(1+2p)(3+2p)\ .
\end{align}
In the above expressions, we have set $B_0=4$ for convenience. Note that $H^\text{(6)}_{1111}$ agrees with the prediction \eqref{higherderivlowestcharge} for $q=1$, modulo the 4-point correlator in the 2-derivative sector given in \eqref{1dsugracorrelator}. In the next subsection, we will explore the extent to which 4d conformal symmetry is realised for higher-derivative corrections. 

\subsection{4d Conformal Symmetry}\label{sec:breaking4dconfsym}

In the free theory, 4d conformal symmetry was realised by acting on the correlators $\langle \cO_{p_1}\cO_{p_2}\cO_{p_3}\cO_{p_4}\rangle$ with a second-order Casimir, while in the 2-derivative sector it was realised directly on the correlators without acting with a Casimir. In section \ref{interp}, it was further observed that the infinite tower of 2-derivative correlators can be assembled into a 4d spin-0 block, which reduces to a scale-invariant amplitude in the flat space limit. Since higher-derivative corrections will give scattering amplitudes that are no longer scale-invariant, this suggests that they can be obtained by acting with Casimirs on 4d conformal blocks. More precisely, we expect that a correlator corresponding to a $2N$-derivative interaction in the effective action should come from acting with a Casimir of order $2N$. In this subsection, we find that this can indeed be realised for a subset of higher-derivative correlators.

Let us consider correlators in the six-derivative sector computed in the previous subsection. Since they are obtained from four-derivative interactions in the effective action (which can have 4d spin-0 or spin-2 as explained in section \ref{sec:higherderivcrossing}), we expect that they can be obtained by acting with a fourth-order Casimir on the holomorphic limit of 4d spin-0 and spin-2 blocks:
\begin{equation}\label{casimirequation}
H_{p_{1}p_{2}p_{3}p_{4}}^{\text{(6)}}|_{\log u}=\Delta_{0}^{(4)}\left(\text{4d spin-0 block}\right)_{p_{1}p_{2}p_{3}p_{4}}^{\text{hol}}+\Delta_{2}^{(4)}\left(\text{4d spin-2 block}\right)_{p_{1}p_{2}p_{3}p_{4}}^{\text{hol}}\ ,
\end{equation}
where the left-hand-side is computed from the effective action in \eqref{fourderivSeff} with $B_0=4$ for convenience, and the Casimirs on the right-hand-side are given by
\begin{equation}\label{fourthordercasimir}
\Delta_{i}^{(4)}=a_{i}+b_{i}\Delta^{(2)}+c_{i}\Delta^{(2)}\mathcal{D}_{x}+d_{i}\Delta^{(2)}\mathcal{D}_{y}\ ,
\end{equation}
where $\Delta^{(2)},\,\mathcal{D}_x,\,\mathcal{D}_y$ were defined in~\eqref{Delta2def} and $\left\{ a_{i},b_{i},c_{i},d_{i}\right\} $ are unfixed coefficients.  See section~\ref{hocs} for a discussion of general higher order Casimirs. We restrict the left-hand-side of \eqref{casimirequation} to the $\ln u$ part of the correlator since this encodes the scattering amplitude in the flat space limit and is therefore the part of the correlator that is expected to have 4d conformal symmetry. In the right-hand-side, we first Taylor expand the 4d conformal blocks in $g_{ij}$ and choose coefficients corresponding to correlators with the desired charges as explained in section \ref{sec:sugra4dconfsym}.

Remarkably, if the coefficient of the ambiguity $C_0=0$ then we find a solution to \eqref{casimirequation} for $p_i=pp11$ and the other correlators related by crossing:
\begin{equation}\label{fourthordercasimirfixed}
\Delta_0^{(4)}=\frac{1}{12}\left(2\,\Delta^{(2)}+5\,\left(\Delta^{(2)}\right)^2\right) \ ,\quad \Delta_2^{(4)}=\frac{1}{90}\left(-2\,\Delta^{(2)}+\left(\Delta^{(2)}\right)^2\right)\ .
\end{equation}
Using these Casimir operators, 4d conformal symmetry can be realised by correlators in the 6-derivative sector with $\{p_i\}=pp11$ and crossing versions. This is highly non-trivial, since neither the Casimir operators nor the 4d blocks preserve crossing symmetry beyond $1\leftrightarrow 2$ exchange, but acting with the former on the latter produces crossing-symmetric correlators. On the other hand, we do not find crossing-symmetric solutions to~\eqref{casimirequation} for other charge configurations. 
This suggests that 4d conformal symmetry is generically broken by higher-derivative corrections. We will see further evidence of this when we compute anomalous dimensions in the next section, and will make further comments about this in the context of $\mathcal{N}=4$ SYM in section \ref{n=4}.

\section{Anomalous Dimensions from Unmixing} \label{anomdim}

In this section we will classify double-trace operators appearing in the conformal partial wave expansion of 4-point correlators which have degenerate quantum numbers at leading order in $1/c$. We then review a procedure for resolving this degeneracy and computing the anomalous dimensions of double-trace operators after unmixing. For correlators in the 2-derivative sector we find that the anomalous dimensions are eigenvalues of a second order conformal Casimir, as anticipated in section \ref{interp}, but for correlators in the 6-derivative sector some anomalous dimensions contain square roots suggesting that 4d conformal symmetry is generically broken in agreement with the analysis of section \ref{sec:breaking4dconfsym}.  

\subsection{Degenerate Operators}

Let us start by considering all long double-trace operators in the OPE of two half-BPS operators, $\cO_{p_1},\cO_{p_2}$. In the free theory the space of operators appearing in the OPE will take the form
\begin{equation}\label{exchangedopsq1q2}
\mathcal{O}_{q_1 q_2}=[\mathcal{O}_{q_1} \partial^{\Delta+1-q_1-q_2}\mathcal{O}_{q_2}]_p\ ,\quad\ q_1\leq q_2\ ,
\end{equation}
where $p_1-q_1=p_2-q_2$ and where  $[\_]_p$ denotes projection onto the $p+1$ dimensional $SU(2)$ representation.
Recall the operators $\mathcal{O}_{q_1}$ and $\mathcal{O}_{q_2}$ have half-integer scaling dimensions $(2q_1-1)/2$ and $(2q_2-1)/2$, labelled by integers $q_1$ and $q_2$, according to the conventions introduced in~\eqref{1dhalfbpsnorm}. Thus the scaling dimension of the exchanged operators $\mathcal{O}_{q_1 q_2}$ is $\Delta$ and the $SU(2)$ charge is $p$ ($p$ can take any value from the set $p=q_1+q_2,q_1+q_2-2,..,q_2-q_1$). There are many operators with the same weight $(\Delta,p)$ which we now enumerate similarly to those in $\cN=4$ SYM in~\cite{Aprile:2018efk}.
 
There are two classes of operators in the spectrum, we call them class $A$ and class $B$. We will also distinguish between the cases where $t=\Delta-p$ odd or $t=\Delta-p$ even. We label the allowed values of pairs of $q_1$ and $q_2$ for class $A$ and class $B$ operators with odd or even $t$ with $(q_1^A,q_2^A)$, $(q_1^{B^\text{to}},q_2^{B^\text{to}})$ and $(q_1^{B^\text{te}},q_2^{B^\text{te}})$ respectively. There are $d=d_A+d_{B^\text{to}}+d_{B^\text{te}}$ operators $\mathcal{O}_{q_1 q_2}$ labelled by sets of pairs  $(q_1,q_2)$, where $d_A$ counts class $A$ and $d_{B^\text{to}}$, $d_{B^\text{te}}$ class $B$ operators with odd and even $t$ respectively. The operators for odd and even $t$ are summarised in Figures~\ref{fig:opsatoddt} and~\ref{fig:opsatevent}, respectively.

For $t$ odd we have both class $A$ and class $B$ operators:
\begin{align}\label{paramoddt}
&q_1^A=1+i_A+r_A\ ,\quad &&q_1^{B^\text{to}}=1+i_{B^\text{to}}+r_{B^\text{to}}\ ,\nonumber \\
&q_2^A=1+i_A+p-r_A\ ,\quad &&q_2^{B^\text{to}}=1+i_{B^\text{to}}+(p-1)-r_{B^\text{to}}\ ,\nonumber \\
&i_A=0,\dots,\frac{t-1}{2}\ ,\quad &&i_{B^\text{to}}=1,\dots,\frac{t-1}{2}\ ,\nonumber \\
&r_A=0,\dots,\frac{\mu_A-1}{2}\ ,\quad &&r_{B^\text{to}}=0,\dots,\frac{\mu_{B^\text{to}}-1}{2}\ ,\end{align}
where 
\begin{equation}
\mu_A=
    \begin{cases}
      p+1\ &p\ \text{even}\\
     p\ &p\ \text{odd}\\
    \end{cases}\ ,\qquad \mu_{B^\text{to}}=
    \begin{cases}
      p-1\ &p\ \text{even}\\
     p\ &p\ \text{odd}
    \end{cases}\ , 
\end{equation}
with $d_A=\tfrac14 (\mu_A+1)(t+1)$ and $d_{B^\text{to}}=\tfrac14 (\mu_{B^\text{to}}+1)(t-1)$\ . Class $A$ operators have even numbers of derivatives while class $B$ operators have odd numbers of derivatives at $t$ odd.

For even $t$, there are only class B operators.
These operators are parametrised as
\begin{align}\label{paramevent}
&q_1^{B^\text{te}}=i_{B^\text{te}}+\lfloor{\tfrac{r_{B^\text{te}}+1}{2}\rfloor}\ ,\nonumber \\
&q_2^{B^\text{te}}=i_{B^\text{te}}+p-\lfloor{\tfrac{r_{B^\text{te}}}{2}\rfloor}\ ,\nonumber \\
&i_{B^\text{te}}=1,\dots,\frac{t}{2}\ ,\nonumber \\
&r_{B^\text{te}}=0,\dots,p-1\ ,
\end{align}
with $d_{B^\text{te}}=\tfrac12\, t\,p$\ . 

\begin{figure}
\begin{subfigure}{16cm}
\centering{\resizebox{9cm}{!}{
\centering \begin{tikzpicture}
\draw[->] (0,0) -- (9,0) node[anchor = south east]{$q_1^A$, $q_1^{B^\text{to}}$};
\draw[->] (0,0) -- (0,10) node[anchor = north west]{$q_2^A$, $q_2^{B^\text{to}}$};

\fill[gray!30] (0.7,3.7) rectangle ++(1.6,2.6);

\fill (1,5) circle[radius=2pt];
\draw[thin] (1,5) -- (2,6);
\fill (2,6) circle[radius=2pt];
\draw[thin] (2,6) -- (3,7);
\fill (3,7) circle[radius=2pt];
\draw[thin] (3,7) -- (4,8);
\fill (4,8) circle[radius=2pt];
\draw[thin] (4,8) -- (5,9);
\fill (5,9) circle[radius=2pt];
\fill (2,4) circle[radius=2pt];
\draw[thin] (2,4) -- (3,5);
\fill (3,5) circle[radius=2pt];
\draw[thin] (3,5) -- (4,6);
\fill (4,6) circle[radius=2pt];
\draw[thin] (4,6) -- (5,7);
\fill (5,7) circle[radius=2pt];
\draw[thin] (5,7) -- (6,8);
\fill (6,8) circle[radius=2pt];
\fill (3,3) circle[radius=2pt];
\draw[thin] (3,3) -- (4,4);
\fill (4,4) circle[radius=2pt];
\draw[thin] (4,4) -- (5,5);
\fill (5,5) circle[radius=2pt];
\draw[thin] (5,5) -- (6,6);
\fill (6,6) circle[radius=2pt];
\draw[thin] (6,6) -- (7,7);
\fill (7,7) circle[radius=2pt];
\draw[thin] (7,7) -- (5,9);
\draw[thin] (1,5) -- (3,3);

\draw[thin,dashed] (2,5) -- (3,6);
\draw[thin,dashed] (3,6) -- (4,7);
\draw[thin,dashed] (4,7) -- (5,8);
\draw[thin,dashed] (2,5) -- (3,4);
\draw[thin,dashed] (3,4) -- (4,5);
\draw[thin,dashed] (4,5) -- (5,6);
\draw[thin,dashed] (5,6) -- (6,7);
\draw[thin,dashed] (5,8) -- (6,7);

\draw[thin] (2,4) -- (2,6);
\draw[thin] (3,3) -- (3,7);
\draw[thin] (4,4) -- (4,8);
\draw[thin] (5,5) -- (5,9);
\draw[thin] (6,6) -- (6,8);

\draw[fill=white, thick] (2,5) circle[radius=2pt];
\draw[fill=white, thick] (3,6) circle[radius=2pt];
\draw[fill=white, thick] (4,7) circle[radius=2pt];
\draw[fill=white, thick] (5,8) circle[radius=2pt];
\draw[fill=white, thick] (3,4) circle[radius=2pt];
\draw[fill=white, thick] (4,5) circle[radius=2pt];
\draw[fill=white, thick] (5,6) circle[radius=2pt];
\draw[fill=white, thick] (6,7) circle[radius=2pt];

\draw[->] (6,9) -- (7,8) node[anchor= south west]{$r_A$, $r_{B^\text{to}}$};
\draw[->] (2,7) -- (3,8) node[anchor= east]{$i_A$, $i_{B^\text{to}}$};

\node at (0.8,5) {A};
\node at (3,2.75) {B};
\node at (7.25,7) {C};
\node at (5,9.3) {D};

\node at (1.75,5) {E};
\node at (3.2,3.75) {F};
\node at (6.3,7) {G};
\node at (5.2,8.3) {H};

\node at (9.5,4.5) {\begin{tabular}{l} $A=\left(1,1+p\right)$ \\ $B=\left(\tfrac12\left(\mu_A+1\right),\tfrac12\left(-\mu_A+2 p+3\right)\right)$ \\ $C=\left(\tfrac12\left(\mu_A+t\right),\tfrac12\left(-\mu_A+t+2 p+2\right)\right)$ \\ $D=\left(\tfrac12\left(t+1\right),\tfrac12\left(t+2 p+1\right)\right)$ \end{tabular}};

\node at (7.5,2) {\begin{tabular}{l} $E=\left(2,1+p\right)$ \\ $F=\left(\tfrac12\left(\mu_{B^\text{to}}+3\right),\tfrac12\left(-\mu_{B^\text{to}}+2 p+3\right)\right)$ \\ $G=\left(\tfrac12\left(\mu_{B^\text{to}}+t\right),\tfrac12\left(-\mu_{B^\text{to}}+t+2 p\right)\right)$ \\ $H=\left(\tfrac12\left(t+1\right),\tfrac12\left(t+2 p-1\right)\right)$ \end{tabular}};

\end{tikzpicture}}}
\caption[Exchanged double-trace operators for odd $t$]{Exchanged operators $\mathcal{O}_{q_1 q_2}$  
for odd $t$ parametrised according to~\eqref{paramoddt}. The black and white nodes denote class $A$ and class $B^\text{to}$ operators, respectively. The only operator acquiring non-zero anomalous dimension in the 2-derivative sector is the one denoted by $A$. The nodes in the grey rectangle correspond to operators which acquire non-zero anomalous dimensions in the six-derivative sector.}\label{fig:opsatoddt}
\end{subfigure}\vspace{.6cm}

\begin{subfigure}{16cm}
\centering{\resizebox{9cm}{!}{
\begin{tikzpicture}
\draw[->] (0,0) -- (9,0) node[anchor = south east]{$q_1^{B^\text{te}}$};
\draw[->] (0,0) -- (0,10) node[anchor = north west]{$q_2^{B^\text{te}}$};

\fill[gray!30] (0.7,4.7) rectangle ++(1.6,0.6);

\fill (1,5) circle[radius=2pt];
\draw[thin] (1,5) -- (2,6);
\fill (2,6) circle[radius=2pt];
\draw[thin] (2,6) -- (3,7);
\fill (3,7) circle[radius=2pt];
\draw[thin] (3,7) -- (4,8);
\fill (4,8) circle[radius=2pt];
\draw[thin] (4,8) -- (5,9);
\fill (5,9) circle[radius=2pt];
\draw[thin] (2,4) -- (1,5);
\fill (2,4) circle[radius=2pt];
\draw[thin] (2,4) -- (3,5);
\fill (3,5) circle[radius=2pt];
\draw[thin] (3,5) -- (4,6);
\fill (4,6) circle[radius=2pt];
\draw[thin] (4,6) -- (5,7);
\fill (5,7) circle[radius=2pt];
\draw[thin] (5,7) -- (6,8);
\fill (6,8) circle[radius=2pt];
\draw[thin] (6,8) -- (5,9);

\draw[thin,dashed] (2,5) -- (3,6);
\draw[thin,dashed] (3,6) -- (4,7);
\draw[thin,dashed] (4,7) -- (5,8);
\draw[thin,dashed] (5,8) -- (6,9);
\draw[thin,dashed] (2,5) -- (3,4);
\draw[thin,dashed] (3,4) -- (4,5);
\draw[thin,dashed] (4,5) -- (5,6);
\draw[thin,dashed] (5,6) -- (6,7);
\draw[thin,dashed] (6,7) -- (7,8);
\draw[thin,dashed] (7,8) -- (6,9);

\draw[fill=white, thick] (2,5) circle[radius=2pt];
\draw[fill=white, thick] (3,6) circle[radius=2pt];
\draw[fill=white, thick] (4,7) circle[radius=2pt];
\draw[fill=white, thick] (5,8) circle[radius=2pt];
\draw[fill=white, thick] (6,9) circle[radius=2pt];
\draw[fill=white, thick] (3,4) circle[radius=2pt];
\draw[fill=white, thick] (4,5) circle[radius=2pt];
\draw[fill=white, thick] (5,6) circle[radius=2pt];
\draw[fill=white, thick] (6,7) circle[radius=2pt];
\draw[fill=white, thick] (7,8) circle[radius=2pt];

\draw[->] (6,10) -- (7,9) node[anchor= south west]{$r_{B^\text{te}}$};
\draw[->] (2,7) -- (3,8) node[anchor= east]{$i_{B^\text{te}}$};

\node at (0.8,5) {I};
\node at (2,3.75) {J};
\node at (6.25,8) {K};
\node at (5,9.3) {L};

\node at (1.75,5) {M};
\node at (3,3.75) {N};
\node at (7.3,8) {O};
\node at (6,9.3) {P};

\node at (8.8,5) {\begin{tabular}{l} $I=\left(1,1+p\right)$ \\ $J=\left(\lfloor{\frac{p+1}{2}\rfloor},p-\lfloor{\frac{p+1}{2}\rfloor}+2\right)$ \\ $K=\left(\lfloor{\frac{p+1}{2}\rfloor}+\frac{t-2}{2},p-\lfloor{\frac{p+1}{2}\rfloor}+\frac{t+2}{2}\right)$ \\ $L=\left(\frac{t}{2},\frac{t}{2}+p\right)$ \end{tabular}};

\node at (7,2) {\begin{tabular}{l} $M=\left(2,1+p\right)$ \\ $N=\left(\lfloor{\frac{p+2}{2}\rfloor},p-\lfloor{\frac{p+2}{2}\rfloor}+3\right)$ \\ $O=\left(\lfloor{\frac{p+2}{2}\rfloor}+\frac{t-2}{2},p-\lfloor{\frac{p+2}{2}\rfloor}+\frac{t+4}{2}\right)$ \\ $P=\left(1+\frac{t}{2},\frac{t}{2}+p\right)$ \end{tabular}};

\end{tikzpicture}}}
\caption[Exchanged double-trace operators for even $t$]{The exchanged operators $\mathcal{O}_{q_1 q_2}$ for even $t$ parametrised according to~\eqref{paramevent}. All operators contributing at even $t$ belong to class B and are split into two groups which do not mix with each other. The black and white nodes denote operators with even and odd $r_{B^\text{te}}$, respectively. The nodes in the grey rectangle correspond to operators which acquire non-zero anomalous dimensions in the six-derivative sector.}\label{fig:opsatevent}
\end{subfigure}
\end{figure}

\subsection{Unmixing}\label{sec:unmixing}

We have seen above that there can be many different double-trace operators with the same quantum numbers $(\Delta,p)$ at leading order in $1/c$. To solve this mixing problem we follow the method developed for $\mathcal{N}=4$ SYM in \cite{Aprile:2017xsp,Drummond:2020dwr}.

First we expand the correlator in terms of long blocks as follows:
\begin{align}\label{Hconformalblockexpansion}
H(x,y)=\frac{x\,y}{x-y}\sum_{\Delta,p}A^{p_i}_{\Delta,p} B^\text{long}_{\Delta,p,p_{12},p_{34}}(x,y)\ ,
\end{align}
where the superconformal blocks are given in~\eqref{1dlongblocks}. The coefficients of the decomposition are given as a sum of squares of OPE coefficients as follows
\begin{equation}\label{OPEcoefficients}
A^{p_i}_{\Delta,p}=\sum_{\cO^{\Delta,p}} C_{p_1 p_2\mathcal{O}} C_{p_3 p_4 \mathcal{O}}\ ,
\end{equation}
where the sum goes over the degenerate operators. Expanding the OPE data in $1/c$ and $a$ (which is the analogue of a string coupling introduced in section \ref{sec:4dscalarSeff}), gives
\begin{align}\label{OPEdataexpansion}
\Delta_{\cO}&=\Delta^{(0)}+\frac{1}{c}\left(\gamma^\text{(2)}+a\,\gamma^\text{(6)}+\dots\right)+\mathcal{O}(c^{-2})\ ,\notag\\
C_{pp\mathcal{O}}&=C_{pp\mathcal{O}}^{(0)}+\left(a\, C_{pp\mathcal{O}}^\text{(6)}+\dots\right)+\cO(c^{-1})\ ,
\end{align}
where the anomalous dimensions $\gamma$ depend on $\left\{ \Delta,p,i,r\right\} $, and the superscripts $\left\{ 0,2,6\right\}$ label the free, 2-derivative, and 6-derivative sectors, respectively. Note that we allow for $\mathcal{O}(a)$ corrections to the OPE coefficients which are not suppressed in $1/c$. Such corrections could in principle arise and were shown to be non-zero in $\mathcal{N}=4$ SYM \cite{Drummond:2020dwr}. 

Plugging \eqref{OPEdataexpansion} into~\eqref{Hconformalblockexpansion} then gives
\begin{align}\label{conformalblockexpansion1/c}
H(x,y)=&\frac{x\,y}{x-y}\sum_{\Delta^{(0)},\,p}\big[A^{{(0)}}_{\Delta^{(0)},\,p}B^\text{long}_{\Delta^{(0)},\,p,p_{12},p_{34}}(x,y)\notag\\
&+ \frac{1}{c}\log u \sum_{\Delta^{(0)},\,p}\left(M^{\text{(2)}}_{\Delta^{(0)},\,p}+a\,M^{\text{(6)}}_{\Delta^{(0)},\,p}+\dots\right)B^\text{long}_{\Delta^{(0)},\,p,p_{12},p_{34}}(x,y)+\dots\big]\ ,
\end{align}
where
\begin{align}\label{OPEcoeffsfreesugraderiv}
A^{(0)}_{\Delta,p}&=\sum_{\cO^{\Delta,p}}C^{(0)}_{p_1 p_2 \cO}C^{(0)}_{p_3 p_4 \cO}\ ,\quad M^{\text{(2)}}_{\Delta,p}=\sum_{\cO^{\Delta,p}} \gamma^\text{(2)}  C^{(0)}_{p_1 p_2 \cO} C^{(0)}_{p_3 p_4 \cO}\ ,\notag\\
 M^{\text{(6)}}_{\Delta,p}&=\sum_{\cO^{\Delta,p}}\left( \gamma^\text{(6)}  C^{(0)}_{p_1 p_2 \cO} C^{(0)}_{p_3 p_4 \cO}+\gamma^\text{(2)}  C^{(0)}_{p_1 p_2 \cO} C^\text{(6)}_{p_3 p_4 \cO}+\gamma^\text{(2)}  C^\text{6}_{p_1 p_2 \cO} C^{(0)}_{p_3 p_4 \cO}\right)\ , \notag \\
0&=\quad \sum_{\cO^{\Delta,p}}C_{p_{1}p_{2}\cO}^{(0)}C_{p_{3}p_{4}\cO}^{(6)}+C_{p_{1}p_{2}\cO}^{(6)}C_{p_{3}p_{4}\cO}^{(0)}\ ,
\end{align}
and we denote $\Delta^{(0)}$ by $\Delta$ for simplicity. The last line follows from the fact that there are no $\mathcal{O}(a)$  corrections to the correlator since in the bulk stringy states only appear on internal lines and are therefore accompanied by Newton's constant to give $\mathcal{O}(a/c)$ corrections. Note that $A^{(0)}$, $M^\text{(2)}$, $M^\text{(6)}$ are determined from the conformal partial wave expansions of correlators in the free, 2-derivative, and 6-derivative sectors, respectively. Given this data, one then solves the above equations for the anomalous dimensions and OPE coefficients $\gamma$ and $C$, respectively. These are referred to as the unmixing equations.

It is convenient to write \eqref{OPEcoeffsfreesugraderiv} in matrix form as follows:
\begin{alignat}{3}\label{unmixingequations}
\mathcal{O}(1):&&\ \hat{A}_{\Delta,p}^{(0)}&=\mathbb{C}^{(0)}\left(\mathbb{C}^{(0)}\right)^T\ ,\notag\\
\mathcal{O}\left(1/c\right):&&\ \hat{M}_{\Delta,p}^\text{(2)}&=\mathbb{C}^{(0)}\hat{\gamma}^\text{(2)}\left(\mathbb{C}^{(0)}\right)^T\ ,\notag\\
\mathcal{O}(a):&&\ 0&=\mathbb{C}^{(0)}\left(\mathbb{C}^\text{(6)}\right)^T+\mathbb{C}^\text{(6)}\left(\mathbb{C}^{(0)}\right)^T\ ,\notag\\
\mathcal{O}\left(a/c\right):&&\ \hat{M}^\text{(6)}_{\Delta,p}&=\mathbb{C}^{(0)}\hat{\gamma}^\text{(6)}\left(\mathbb{C}^{(0)}\right)^T+\mathbb{C}^{(0)}\hat{\gamma}^\text{(2)}\left(\mathbb{C}^\text{(6)}\right)^T+\mathbb{C}^{\text{(6)}}\hat{\gamma}^\text{(2)}\left(\mathbb{C}^{(0)}\right)^T\ .
\end{alignat} 
In more detail, the matrices are $\left((d_A+d_{B^\text{to}})\times (d_A+d_{B^\text{to}})\right)$ and $(d_{B^\text{te}}\times d_{B^\text{te}})$ for odd and even $t$, respectively. Note that $\hat{\gamma}$ is a diagonal matrix of anomalous dimensions, $\hat{A}^{(0)}$ is a diagonal matrix, and $\hat{M}^\text{(2)}$ and  $\hat{M}^\text{(6)}$ are symmetric matrices. When $t$ is odd, the matrices are block-diagonal, so class $A$ and $B$ operators can be treated independently, and similarly when $t$ is even operators with $r_{B^\text{te}}$ even or odd can be treated independently.

In practice, it is useful to define a matrix $\tilde{c}$ such that
\begin{align}\label{unmixingmatrixform0}
\tilde{c}\tilde{c}^T=\text{Id}\ ,\quad \mathbb{C}^{(0)}=\left(\hat{A}^{(0)}\right)^{\tfrac12}\cdot \tilde{c}\ .
\end{align}
The unmixing equations in the 2-derivative sector then become:
\begin{align}\label{unmixingmatrixform}
\tilde{c}\cdot\hat{\gamma}^\text{(2)}\cdot\tilde{c}^T=\left(\hat{A}^{(0)}\right)^{-\frac12}\cdot\hat{M}^\text{(2)}\cdot\left(\hat{A}^{(0)}\right)^{-\frac12}\ .
\end{align}
From this we see that the columns of $\tilde{c}$ are eigenvectors of the matrix $\left(\hat{A}^{(0)}\right)^{-\frac12}\cdot\hat{M}^\text{(2)}\cdot\left(\hat{A}^{(0)}\right)^{-\frac12}$ and the corresponding eigenvalues are the anomalous dimensions. The anomalous dimensions can also be computed directly as the eigenvalues of the matrix $\hat{M}^{\text{(2)}}\cdot\left(\hat{A}^{(0)}\right)^{-1}$. Similarly, the anomalous dimensions in the 6-derivative sector are the $\cO(a)$ eigenvalues of $\left(\hat{M}^\text{(2)}+a\hat{M}^\text{6-deriv}\right)\cdot\left(\hat{A}^{(0)}\right)^{-1}$.   

Let us illustrate the method with a few simple examples in the 2-derivative sector. For the simplest case $\left(\Delta,p\right)=\left(1,0\right)$ there is only one exchanged operator of the form~\eqref{exchangedopsq1q2}: $\cO_{1} \cO_{1}$. Performing the conformal block expansion of the correlator $H^\text{(2)}_{1111}$ in \eqref{1dsugracorrelator} and comparing to \eqref{conformalblockexpansion1/c}, we find that the coefficients of the expansion for $\left(\Delta,p\right)=\left(1,0\right)$ are
\begin{align}\label{freetheorysugra10}
A^{(0)}_{1111}(1,0)=1\ ,\quad M^\text{(2)}_{1111}(1,0)=2.
\end{align}
The unmixing equations in \eqref{unmixingequations} are then simply
\begin{equation}
A^{(0)}_{1111}(1,0)=\left(C^{(0)}_{1,0}\right)^2\ ,\quad M^\text{(2)}_{1111}(1,0)=\gamma^\text{(2)}_{1,0}\left(C^{(0)}_{1,0}\right)^2\ .
\end{equation}
Solving these equations then gives
\begin{equation}\label{unmixedcoeffssugra10}
\gamma^\text{(2)}_{1,0}=2\ ,\quad C^{(0)}_{1,0}=1\ .
\end{equation}

The next simplest example is $\left(\Delta,p\right)=\left(3,0\right)$. In this case, there are two possible exchanged operators, $\cO_{2}\cO_{2}$ and $\cO_{1} \partial^2 \cO_{1}$, and we have
  \begin{align}\label{freetheorysugra30}
\hat{A}^{(0)}_{3,0}=
\begin{pmatrix}
A^{(0)}_{1111} & 0 \\
0 & A^{(0)}_{2222}
\end{pmatrix}_{(3,0)}=
\begin{pmatrix}
\frac{1}{10} & 0 \\
0 & \frac{1}{18}
\end{pmatrix}\ ,\  \hat{M}^\text{(2)}_{3,0}=
\begin{pmatrix}
M^\text{(2)}_{1111} & M^\text{(2)}_{1122} \\
M^\text{(2)}_{1122} & M^\text{(2)}_{2222}
\end{pmatrix}_{(3,0)}=\begin{pmatrix}
\frac{1}{5} & \frac{1}{3} \\
\frac{1}{3} & \frac{5}{9}
\end{pmatrix}\ .
\end{align}
Solving the unmixing equations in \eqref{unmixingequations}:
\begin{align}\label{unmixedcoeffssugra30}
\hat{\gamma}^\text{(2)}_{3,0}=\begin{pmatrix}
12 & 0 \\
0 & 0
\end{pmatrix}\ ,\quad \mathbb{C}^{(0)}_{3,0}=\begin{pmatrix}
\frac{1}{2\sqrt{15}} & \frac{1}{2\sqrt{3}} \\
\frac{\sqrt{5}}{6\sqrt{3}} & -\frac{1}{6\sqrt{3}}
\end{pmatrix}\ .
\end{align}
Note that there is only one non-zero anomalous dimension. This structure extends to general $\Delta$ and $p$ in the 2-derivative sector: at each weight there is only one non-zero anomalous dimension, which is a rational number. In the 6-derivative sector all the anomalous dimensions will generically be non-zero and contain square roots. We will provide an explanation for this in the next subsection. See Appendix \ref{unmixingexamples} for further examples of unmixing.

\subsection{Anomalous Dimensions}

We will now apply the unmixing procedure described in the previous subsection to compute anomalous dimensions of double trace operators contributing to the conformal partial wave expansion of 4-point correlators in the 2-derivative and 6-derivative sectors. We obtain general formulas for the anomalous dimensions by working out many examples.

Let us start with the 2-derivative sector. In that case we find that only class $A$ operators acquire anomalous dimensions while class $B$ operators decouple. In more detail, we find that only one operator acquires a non-zero anomalous dimension for each $\left(\Delta,p\right)$, and the value is given by
 \begin{equation}\label{delta2eigenvalue}
 \gamma_{\Delta,p}^\text{(2)}=\delta^{(2)}=(\Delta-p)(\Delta+1+p). 
 \end{equation}
We recognize this to be the eigenvalue of the Casimir $\Delta^{(2)}$ in \eqref{Delta2def} acting on superconformal blocks:
\begin{equation}
\Delta^{(2)}\left(\frac{x\,y}{x-y}B^\text{long}_{\Delta,p,p_{12},p_{34}}\right)=\delta^{(2)}\left(\frac{x\,y}{x-y}B^\text{long}_{\Delta,p,p_{12},p_{34}}\right)\ .
\end{equation}
This was anticipated from 4d conformal symmetry in~\eqref{predictiondelta2}. The fact that only one operator acquires a non-zero anomalous dimension for each $\left(\Delta,p\right)$ can be understood from the following formula relating 4d spin to 1d quantum numbers:
\begin{align}\label{4dspin}
l_{4d}=2\left(i_A+r_A+i_{B^\text{to}}+i_{B^\text{te}}+r_{B^\text{te}}-\frac{1+(-1)^{r_{B^\text{te}}+1}}{2}\right)\ ,
\end{align}
which has an analogous form in $\mathcal{N}=4$ SYM \cite{Drummond:2020dwr}. In particular, recalling that correlators in the 2-derivative sector come from the 4d spin-0 block (as explained in section \ref{interp}) and setting $l_{4d}=0$, we see that class $B$ operators cannot contribute since $i_B \geq 1$ and only the class $A$ operator with $i_A=r_A=0$ can contribute.

We can similarly see that  both class $A$ and class $B$ operators acquire anomalous dimensions in the six derivative sector, since the corresponding correlators can be expanded in terms of 4d blocks with spin-0 and spin-2, as explained in section \ref{sec:higherderivcrossing}. It is useful to separately consider the cases where $t=\Delta-p$ is odd or even. In the results below we have set the coefficient $B_0=4$ in \eqref{fourderivSeff} for notational simplicity. One can obtain anomalous dimensions for general $B_0$ by inserting a factor of $B_0/4$ in all terms that are not multiplied by $C_0$.

\subsubsection*{odd $t$}

As explained above, there are two possibilities: $l_{4d}=0,2$. The condition $l_{4d}=0$ can only be satisfied by class $A$ operators with quantum numbers $i_A=r_A=0$ and in this case the anomalous dimensions are 
   \begin{equation}\label{anomdimoddti0r0}
\left(\gamma^A\right)_{0,0}^{\Delta,p}=\delta^{(2)}\left(\frac{2}{9}\,(1-6\,C_0)+\frac{1}{18}\,(1+12\,C_0)\,\delta^{(2)}+\frac{5}{12}\,\left(\delta^{(2)}\right)^2+\frac{1}{9}\,(1+12\,C_0)\,\delta^{(y)}\right)\ ,
\end{equation}
where $\delta^{(2)}$ and $\delta^{(y)}$ are the eigenvalues of $\Delta^{(2)}$ and $\mathcal{D}_y$ in~\eqref{Delta2def}, respectively:
\begin{equation}
\delta^{(2)}=(\Delta-p)(\Delta+p+1)\ ,\quad \delta^{(y)}=p(p+1)\ .
\end{equation}
For $l_{4d}=2$, there are three possible non-zero anomalous dimensions: $\left(i_{A},r_{A}\right)\in\left\{ \left(1,0\right),\left(0,1\right)\right\} $ for class $A$, and $i_{B^\text{to}}=1$ for class $B$, and the corresponding anomalous dimensions are
\begin{align}
\left(\gamma^A\right)_{1,0}^{\Delta,p}=&\frac{1}{180}\,\delta^{(2)}\Big(4-2\,\delta^{(2)}+3\,\left(\delta^{(2)}\right)^2-4\,\delta^{(y)}\notag\\
&-4\,\sqrt{8\,p^3+4\,p^4-4\,p^2 \Delta(1+\Delta)-4\,p\,(1+\Delta+\Delta^2)+(-1+2\,\Delta+2\,\Delta^2)^2}\Big)\ ,\label{anomdimoddti1r0}\\
\left(\gamma^A\right)_{0,1}^{\Delta,p}=&\frac{1}{180}\,\delta^{(2)}\Big(4-2\,\delta^{(2)}+3\,\left(\delta^{(2)}\right)^2-4\,\delta^{(y)}\notag\\
&+4\,\sqrt{8\,p^3+4\,p^4-4\,p^2 \Delta(1+\Delta)-4\,p\,(1+\Delta+\Delta^2)+(-1+2\,\Delta+2\,\Delta^2)^2}\Big)\ ,\label{anomdimoddti0r1}\\
\left(\gamma^{B^\text{to}}\right)_1^{\Delta,p}=&\frac{1}{60}\,\delta^{(2)}\left(-2\,\delta^{(2)}+\left(\delta^{(2)}\right)^2-4\,\delta^{(y)}\right)\ .\label{anomdimoddtiB1}
\end{align}
Unlike in the 2-derivative sector, all operators have non-zero anomalous dimensions and the latter contain square roots, which signals the breaking of 4d conformal symmetry. This is consistent with the conclusions in section \ref{sec:breaking4dconfsym}.

\subsubsection*{even $t$}

For even $t$, operators only start to contribute for $l_{4d}=2$ and we get two different non-zero anomalous dimensions for quantum numbers $\left(i_{B^{\text{te}}},r_{B^{\text{te}}}\right)\in\left\{ \left(1,0\right),(1,1)\right\} $:
\begin{equation}\label{anomdimeventr0}
\left(\gamma^{B^\text{te}}\right)_{1,0}^{\Delta,p}=\frac{1}{60}\left(\delta^{(2)}\right)^2\left(\delta^{(2)}-2\right)\ ,
\end{equation}
\begin{equation}\label{anomdimeventr1}
\left(\gamma^{B^\text{te}}\right)_{1,1}^{\Delta,p}=\frac{1}{60}\left(\delta^{(2)}\right)^2\left(\delta^{(2)}+2\right)\ .
\end{equation}
Note that \eqref{anomdimeventr0} is proportional to the eigenvalue of $\Delta_\text{2}^{(4)}$ in~\eqref{fourthordercasimirfixed}.

\section{Comments on $\mathcal{N}=4$ SYM} \label{n=4}

The main focus of this paper has been on 1d $\cN=4$ SCFT, its relation to a bulk theory on AdS${}_2\times$S$^2$ and in particular the investigation of a hidden 4d symmetry. The primary motivation for this came from the 4d $\cN=4$ SCFT context, related to string theory in AdS${}_5 \times$S$^5$, which was shown to possess a  hidden 10d symmetry~\cite{Caron-Huot:2018kta}.
This 10d symmetry has been further investigated in a number of works (see e.g.
\cite{Aprile:2020luw,Abl:2020dbx,Aprile:2020mus,Caron-Huot:2021usw}). It was also generalised to AdS${}_3 \times$S$^3$ in~\cite{Rastelli:2019gtj,Giusto:2020neo,Aprile:2021mvq,Wen:2021lio}.
Thus we can view the AdS${}_2 \times$S$^2$ setting here as  a toy model for the AdS${}_5 \times$S$^5$ case. 
In this section we then draw out some of the lessons learned in AdS${}_2 \times$S$^2$ and apply them to $\cN=4$ SYM.

We will only very briefly review the $\cN=4$, AdS${}_5 \times$S$^5$ story here, and will do so mainly by comparison to the 1d story detailed in earlier parts of the paper. Indeed the entire  formalism we have used makes the generalisation particularly natural, needing minimal modification.
The SCFT ($\cN=4$ SYM) naturally lives on analytic superspace~\cite{Galperin:1984av,Howe:1995md,Doobary:2015gia}
 which has the same matrix form as in the 1d case~\eqref{coords} but where each term $x,y,\theta,\bar \theta$ becomes itself  a $2\times 2$ matrix. The half BPS superfield of~\eqref{Ldescendant} thus has more component fields in its expansion -- since there are now four $\theta$s and four $\bar \theta$s -- but we will still focus on just two important component fields: the highest weight state $\cO_p(x,y)$ and the highest term in the $\theta$ (only) expansion,  $L_p(x,y)$:
\begin{align}
	{\bf{O}}_p(x,y) = \cO_p(x,y)+..+\theta^4 L_p(x,y)+.. +\bar{\theta}^4\bar{L}_p(x,y)+.. \ .
\end{align}
Here $\cO_p$ has scaling dimension $p$ and $SU(4)$ rep $[0,p,0]$ whereas $L_p$ has dimension $p+2$ and  $SU(4)$ rep $[0,p-2,0]$.

As in the 1d case (see \eqref{Lgen}, \eqref{ogen})  we define a single object which generates all (appropriately normalised) half BPS operators $\cO_p$ and their descendants $L_p$:
\begin{align}
	\cO= \sum_{p=2}^\infty \cO_p \qquad \qquad 	L= \sum_{p=2}^\infty L_p\ .
\end{align}
Then the  10d symmetry of free and $1/c$ (tree level supergravity) discovered in~\cite{Caron-Huot:2018kta} takes the form (compare this with the 1d SCFT equivalent cases with manifest 4d symmetry~\eqref{LLLbLb4d},\eqref{4d2deriv})
\begin{align}\label{10dfree}
	\langle LL\bar L \bar L \rangle_{c^0}  &=  \left[\frac1{x_{14}^8x_{23}^8}+\frac{1}{x_{13}^8x_{24}^8}\right]_{10d},\\
	\langle \cO \cO \cO \cO\rangle_{1/c}&= \cI_{\cN=4} \times \left[\frac{\bar D_{2422}}{x_{13}^8x_{24}^8}\right]_{10d},\ 
\end{align}
where we are neglecting overall numerical coefficients,
\begin{align}
	\cI_{\cN=4}& = (x-y)(x-\bar y)(\bar x-y)(\bar x-\bar y)x_{13}^4x_{24}^4y_{13}^4y_{24}^4,
\end{align}
and 
\begin{align}
	\frac{x_{12}^2x_{34}^2}{x_{13}^2x_{24}^2}&=x\bar x \qquad \frac{x_{14}^2x_{23}^2}{x_{13}^2x_{24}^2}=(1-x)(1-\bar x)
	\notag\\
	\frac{y_{12}^2y_{34}^2}{y_{13}^2y_{24}^2}&=y\bar y \qquad \frac{y_{14}^2y_{23}^2}{y_{13}^2y_{24}^2}=(1-y)(1-\bar y)\ .
\end{align}

Now similarly to the 1d case (see~\eqref{descendentcorrelator1}), the correlator of descendants is related to the correlator of primaries by a Casimir acting at points 1 and 2. This time the Casimir is  eighth order~\cite{Drummond:2006by,Caron-Huot:2018kta} 
\begin{align}
	\label{descendentcorrelator}
	\langle L_{p_1}L_{p_2}\bar L_{p_3} \bar L_{p_4} \rangle={\mathcal I}_{\cN=4}^{-1}\,\mathcal{C}^{(8);SU(2,2|4)}_{1,2} \langle \cO_{p_1}\cO_{p_2}\cO_{p_3}\cO_{p_4} \rangle\ ,
\end{align}
where 
${\mathcal C}_{1,2}^{(8);SU(2,2|4)}$ denotes an eighth order superconformal Casimir defined explicitly as:
\begin{align}\label{Delta8def}
	{\mathcal C}_{1,2}^{(8);SU(2,2|4)}&= \hat P_{p_i} \left( \frac{(x{-}y)(x{-}\bar y)(\bar x{-}y)(\bar x{-}\bar y)}{x \bar x y \bar y (x{-}\bar x)(y {-}\bar y)}\right)     \Delta^{(8)}   \left(\frac{x \bar x y \bar y (x{-}\bar x)(y {-}\bar y)}{(x{-}y)(x{-}\bar y)(\bar x{-}y)(\bar x{-}\bar y)}\right)    \hat P_{p_i}^{-1}\ , \notag \\
	\Delta^{(8)} &= \left({\mathcal D}^+_x - {\mathcal D}^-_y\right) \left({\mathcal D}^+_{\bar x} - {\mathcal D}^-_y\right)\left({\mathcal D}^+_x - {\mathcal D}^-_{\bar y}\right)\left({\mathcal D}^+_{\bar x} - {\mathcal D}^-_{\bar y}\right)\ ,
\end{align}
with $ {\mathcal D}^\pm_{z}$ defined in~(\ref{Delta2def}c) and $\hat P_{p_i}$ in~\eqref{prefactor} (but with now $g_{ij}=y_{ij}^2/x_{ij}^2$). 
The superconformal Casimirs are related to Casimirs for the maximal bosonic subgroup via conjugation with $\cI_{\cN=4}$ as discussed in~\eqref{superctoboscnis4}:
 \begin{align}\label{casbos}
	\mathcal{C}^{SU(2,2|4)}_{1,2} = \left(\frac{(x{-}y)(x{-}\bar y)(\bar x{-}y)(\bar x{-}\bar y)}{(x \bar x y \bar y)^2}\right) \, \mathcal{C}^{SO(1,5)\times SO(6)}_{1,2} \, \left(\frac{(x \bar x y \bar y)^2}{(x{-}y)(x{-}\bar y)(\bar x{-}y)(\bar x{-}\bar y)}\right)\ .
\end{align}

Putting all this together we see that for the descendant correlator at order $1/c$ (i.e. the supergravity approximation) the 10d symmetry manifests as a $SO(1,5)\times SO(6)$ Casimir acting on a 10d symmetric function (again similarly to in 1d - compare with~\eqref{lllblb2})
\begin{align}\label{lllblb210d}
	\langle LL\bar L\bar L\rangle_{1/c} =\mathcal{C}^{(8);SO(1,5)\times SO(6)}_{1,2} \, \left[\frac{\bar D_{2422}}{x_{13}^8x_{24}^8}\right]_{10d}\ .
\end{align}
We now examine whether higher-derivative corrections can be written as higher-order Casimirs acting on 10d functions. We find that the answer is yes for the $\alpha'^3$ correction, but for higher corrections things are not so simple. 

\subsection{ 10d conformal symmetry at $O(\alpha'^3)$}

Recently the 4-point correlators of all $1/2$-BPS operators at $O(\alpha'^3/c)$ have been obtained in Mellin space~\cite{Goncalves:2014ffa,Alday:2018pdi,Drummond:2019odu} and translated to position space~\cite{Abl:2020dbx}. We now claim that these $\alpha'^3$ corrections can be written as a fourteenth order Casimir of the maximal bosonic subgroup acting on a 10 dimensional function:
\begin{align}\label{preamp}
	\left\langle LL \bar L \bar L \right\rangle|_{\alpha'^3/c}&=  \mathcal{C}^{(8+6);SO(1,5)\times SO(6)}_{1,2} \left[\frac{\bar D_{4411}( u,v)}{x_{13}^8x_{24}^8} \right]_{10d}\ .
\end{align}
where the fourteenth order Casimir factors $\mathcal{C}^{(8+6)}= \mathcal{C}^{(8)}\mathcal{C}^{(6)}$ into the above eighth order Casimir $\mathcal{C}^{(8)}$ times a sixth order Casimir $\mathcal{C}^{(6)}$ which is defined similarly to~\eqref{Delta8def} but in terms of
\begin{align} \label{delta6}
	\Delta^{(6)} &= -\tfrac18 \left( \cD_x^+{+}\cD_{\bar x}^+{-}\cD_y^-{-}\cD_{\bar y}^-\right)^3+\left( (\cD_x^+)^2{+}(\cD_{\bar x}^+)^2{-}(\cD_y^-)^2{-}(\cD_{\bar y}^-)^2\right)-\tfrac32 \left( \cD_x^+{+}\cD_{\bar x}^+{-}\cD_y^-{-}\cD_{\bar y}^-\right)\ .
\end{align}

The form~\eqref{preamp} follows  naturally from  the 10d structure of the free theory~\eqref{10dfree} and the supergravity correction~\eqref{lllblb210d} together with the hint that dimensional analysis of the corresponding term in the effective action translates into a Casimir of a certain order (here sixth order beyond supergravity due to the dimension six coefficient $\alpha'^3$). Further just as the supergravity correction can be expanded in 10d blocks~\cite{Caron-Huot:2018kta} so can the $\alpha'^3$ correction. On the other hand, the $\alpha'^3$ correction arises from an effective  10d AdS${}_5\times$S$^5$ scalar contact Witten diagram~\cite{Abl:2020dbx}, and thus  the classic arguments of~\cite{Heemskerk:2009pn} lifted to the 10d situation imply that only a 10d spin-0 block should appear. Indeed the $\log u$ part of $\bar D_{4411}$ is precisely the 10d spin-0 block, just as the 1d CFT correlators in the two-derivative sector arise from a 4d spin-0 block~\eqref{spin0block1d}.

The precise form of the sixth order Casimir~\eqref{delta6} was found by taking an arbitrary linear combination of the Casimirs of order six and below (or more accurately Casimirs of order 14 down to 8 containing a $\Delta^{(8)}$ factor using the basis in~\eqref{Deltaababbb}). We then evaluated the  action of these operators  on the  $\log u$ part of the 10d function $\bar D_{4411}$ expanded into 4d superblocks. In fact, since the 4d blocks are eigenfunctions of the Casimirs and the eigenvalues are explicitly known~\cite{fh} this becomes simple algebra. The resulting anomalous dimensions were then compared with the results for these anomalous dimensions found in~\cite{Drummond:2020dwr}. Matching the results gave the above unique Casimir.

The result~\eqref{preamp} can instead be written in terms of the correlator of primaries via~\eqref{descendentcorrelator}
\begin{align}\label{preamp2}
	\langle \cO \cO \cO \cO \rangle|_{\alpha'^3/c}&=  \cI_{\cN=4} \times \mathcal{C}^{(6);SO(1,5)\times SO(6)}_{1,2} \left[\frac{\bar D_{4411}( u,v)}{x_{13}^8x_{24}^8} \right]_{10d}\ .
\end{align}
One can also check explicitly that the component correlators agree with known results, for example for the lowest component term one can check that
\begin{align}
	\frac{x \bar x y \bar y}{(x-\bar x)(y -\bar y)} \Delta^{(6)}\frac{(x-\bar x)(y -\bar y)}{x \bar x y \bar y} \left( u^4 \bar D_{4411}\right)=  60 \bar D_{4444}\ .
\end{align}
This then  reproduces the result of~\cite{Goncalves:2014ffa} for this correction. One can also verify higher charge correlators by comparing with the results in Mellin space derived in~\cite{Alday:2018pdi,Drummond:2020dwr} or in position space derived in~\cite{Abl:2020dbx}.

\subsection{ 10d conformal symmetry at $O(\alpha'^5)$} \label{alpha5}

We now wish to see if the above structure generalises to $O(\alpha'^5)$. Similarly to the 6-derivative sector in AdS$_2\times$S$^2$, we will find that it is not possible to reproduce the predictions of the effective action by acting with Casimirs on 10d conformal blocks except for a certain infinite class of operators. 
 We then consider an alternative approach by constructing an integrand with 10d conformal symmetry from the generalised Witten diagrams of the effective action. Remarkably, here we do find a unique candidate possessing 10d symmetry at the level of the integrand. The combination of terms in the effective action which possess this integrand-level symmetry do not agree with the predictions arising from localisation results~\cite{Binder:2019jwn} in $\cN=4$ SYM however. These results do provide an alternative point of view on the breaking of 10d conformal symmetry by $\alpha'^5$ corrections, which was first suggested by the structure of anomalous dimensions obtained after unmixing \cite{Drummond:2020dwr}, and will hopefully lead to a more systematic understanding of how this symmetry is broken.

\subsubsection*{Conformal Functions from 10d Casimirs}

Let us then see if all $1/2$-BPS correlators packaged together can be written as Casimirs acting on 10d conformal functions at $O(\alpha'^5)$. More concretely, since the effective action at $O(\alpha'^5)$ contains four derivatives (see~\cite{Abl:2020dbx}), following the general arguments of \cite{Heemskerk:2009pn} we might expect the 10d function to be related to 10d spin-0 and spin-2 blocks (just as the $O(\alpha'^3)$ correction was related to a 10d spin-0 block). Furthermore, since the $\alpha'^5$ corrections correspond to 10-derivative corrections to supergravity, we expect to act on these blocks with Casimirs of order $10+8=18$. We therefore propose the following structure (similar to the one proposed in 1d~\eqref{casimirequation}):
\begin{align}\label{preamp10d} 
	\left\langle LL \bar L \bar L \right\rangle|_{\alpha'^5/c}|_{\log u}&=  \mathcal{C}^{(8+10)}_{1,2} \left[ \frac{\text{block}_2^{d=10}}{X_{12}^4X_{34}^4}\right]_{10d}+ \mathcal{C}'^{(8+10)}_{1,2} \left[ \frac{\text{block}_0^{d=10}}{X_{12}^4X_{34}^4}\right]_{10d}\ ,
\end{align}
where the 10d blocks can be found in~\cite{Caron-Huot:2018kta}.

Interestingly we find a unique and remarkably simple solution which correctly reproduces all the $1/2$-BPS correlators of the form $\langle 22pp \rangle $ arising from the simplest four-derivative effective action $\int d^{10} x (\nabla \phi.\nabla \phi)^2$ (called $S^\text{main}$ in~(\ref{effact4deriv}a)).
The solution is
\begin{align}\label{proposal}
	\left\langle LL \bar L \bar L \right\rangle|_{\text{main}}|_{\log u}&= \mathcal{C}^{(8)}_{1,2}\mathcal{C}^{(8)}_{1,2} (2-\tfrac12\mathcal{C}^{(2)}_{1,2})\left( \frac{\text{block}_2^{d=10}+ 28 \text{block}_0^{d=10}}{X_{12}^4X_{34}^4}\right) -72 \,\mathcal{C}^{(8)}_{1,2}\mathcal{C}^{(8)}_{1,2}\left( \frac{\text{block}_0^{d=10}}{X_{12}^4X_{34}^4}\right)\ ,
\end{align}
and we note that all the high order Casimirs reduce to products of the eighth order Casimir~\eqref{Delta8def} together with the quadratic Casimir $\mathcal{C}^{(2)}_{1,2}$.\footnote{The quadratic Casimir is defined in  a similar way to~\eqref{Delta8def} in terms of $\Delta^{(2)}=\cD^+_x+\cD^+_{\bar x}-\cD^-_y-\cD_{\bar y}^-$.} However it does not correctly give the more general half BPS operators $\langle p_1 p_2 p_3 p_4 \rangle$ resulting from $S_{\text{main}}$.
This is in complete analogy with what happens in the 1d case (see section~\ref{sec:breaking4dconfsym} in particular the last paragraph) where as soon as derivatives first appear in the effective action, the 4d conformal symmetry is broken.  Furthermore there is no solution of the form~\eqref{preamp10d} which yields the $\alpha'^5/c$ correlators predicted by localisation in $\cN=4$ SYM~\cite{Binder:2019jwn}, even for the case of $\langle 22pp \rangle$ correlators. It therefore appears that breaking of 10d conformal symmetry is related to the presence of derivatives in the effective action.

\subsubsection*{A 10d Conformal Integrand}

This leads one to ask whether instead the higher conformal symmetry might remain at the level of the {\em integrand} of the sum of generalised Witten diagrams derived from the effective action rather than explicitly at the functional level via Casimir operators.%
\footnote{We thank Simon Caron-Huot and Frank Coronado for suggesting this possibility.} 

It turns out that this is the case for a specific four-derivative effective action, although this action  does not apparently coincide with the actual one predicted by the localisation results of~\cite{Binder:2019jwn}. We will write the result for a SCFT with general dimension $d$, corresponding to higher dimensional conformal group $SO(2,2d+2)$. For $\cN=4$ SYM we have $d=4$ whereas for the 1d CFT we have $d=1$. In detail, we find that the following higher dimensional effective Lagrangian:
\begin{align}\label{Leff}
L_{\text{conf}}=&\, \Big(2d^2\nabla_{\mu}\nabla^{2}\nabla_{\mu}\phi\phi^{3} -2d(1+d)	\nabla^{2}\nabla_{\mu}\phi\nabla_{\mu}\phi\phi^{2}-2d^2	\nabla_\nu\nabla_{\mu}\phi\nabla_{\nu}\nabla_\mu\phi\phi^{2}\notag \\
&+4d(1+d)	\nabla_\nu\nabla_{\mu}\phi\nabla_{\mu}\phi\nabla_\nu\phi\phi-(1+d)(1+2d)	\nabla_\mu\phi\nabla_{\mu}\phi\nabla_{\nu}\phi\nabla_\nu\phi\Big)\ ,
\end{align}
has a corresponding Witten diagram expression which has integrand-level 10d conformal symmetry.

Replacing each scalar in \eqref{Leff} by the higher dimensional bulk to boundary propagator~\eqref{bulktoboundary10d} and summing over permutations yields the Witten diagram integrand
\begin{align}\label{integrand10d}
\tfrac16 d^2(1+d)	\sum_{S_4}\left(\frac{\frac{1+d}4 
\bP_1^2 \bP_2^2 
\bX_{34}^2-\frac{1+d}2\bP_1^2 \bP_2 \bP_3 
	\bX_{34} \bX_{24}+\frac{1+2d}8 \bP_1 \bP_2 \bP_3 \bP_4 \bX_{34} 
	\bX_{12}-\frac18 \bP_1^2 \bP_2^2 \bP_3^2 \bP_4^2}{\bP_1^6 
	\bP_2^6\bP_3^6\bP_4^6}\right),
\end{align}
where $\bX_{ij}:= X_{i}.X_j+Y_{i}.Y_{j}$, $\bP_i:=P_i+Q_i=\hat X.X_i+\hat 
Y.Y_i$\ .
We see that this 
can be written entirely in terms of higher dimensional  propagators $P+Q=\hat X.X_i+\hat 
Y.Y_i$ and $SO(2,2d+2)$ invariants $X_{i}.X_j+Y_{i}.Y_{j}$. Furthermore it has uniform weight -4 in each of the projective coordinates $(X_i,Y_i)$. Thus the integrand possesses 10d conformal symmetry.
Compare this for example with the Witten diagrams arising from the simplest forms of the four-derivative Lagrangians in~\eqref{4derivmaincontactdiagram} and~\eqref{4derivambcontactterm} where the numerators can be seen to destroy the higher dimensional conformal symmetry even at the integrand level. 
This is  reminiscent of the integrand level 10d conformal symmetry of perturbative half BPS 4-point functions~\cite{Chicherin:2015edu} noticed in~\cite{Caron-Huot:2021usw}. 

Now the integration of this integrand is over AdS$_5 \times$S$^5$ rather than the Weyl equivalent 10d flat space, and so the breaking of the symmetry must arise from this measure. In the case of $\alpha'^3$ given by a  $\phi^4$ interaction we found that the corresponding integral possesses 10d conformal symmetry, but only after acting with a  sixth order Casimir~\eqref{preamp10d}.  This Casimir then presumably arises as a consequence of performing the Weyl transformation from $AdS_5\times S^5$ to flat space. In the current case the Weyl transformation is no longer so simple however (indeed notice that the weights in the integration variables $\hat X,\hat Y$ is non uniform in~\eqref{integrand10d}) and a  Casimir acting on a 10d function no longer works.

Note that to obtain the Lagrangian in~\eqref{Leff} we wrote down an ansatz involving all possible 4-derivative Lagrangians (so all possible ways of acting with 4 non commuting derivatives on 4 scalars) and imposed that the result be a function of $P_i,Q_i$ through the combinations  $\bP_i:=P_i+Q_i$ only. Interestingly, the result was then automatically a function of $X_{i}.X_j,Y_{i}.Y_{j}$ through the combination   $\bX_{ij}:= X_{i}.X_j+Y_{i}.Y_{j}$ only, without us having to impose this condition directly. Using integration-by-parts the effective Lagrangian possessing the integrand level symmetry~\eqref{Leff} can be written as a specific linear combination of $L_{\text{main}}$ and $L_{\text{amb}}$~\eqref{effact4deriv} as
\begin{align}
	L_{\text{conf}} \, \sim\, (1{+}3d)\Big(
-(1{+}2d)\nabla_\mu\phi\nabla_{\mu}\phi\nabla_{\nu}\phi\nabla_\nu\phi-2d\nabla^2\nabla_\mu\phi\nabla_{\mu}\phi\phi\phi \Big)\ .
\end{align}
Putting  $d=4$ this combination is however not the one predicted by the localisation results of~\cite{Binder:2019jwn} (compare with the combination in~\cite{Abl:2020dbx} eq(71)). Thus we see that although there is a unique four derivative effective action which implies an integrand with 10d conformal symmetry, it is apparently not the one chosen by $\cN=4$ SYM.

\section{Conclusion} \label{conclusion}

In this paper we explored two complementary approaches for computing correlators of hypermultiplets in AdS$_2 \times$S$^2$. The first makes use of 4d conformal symmetry in the free and 2-derivative sectors of the theory, which makes it possible to deduce 4-point correlators of protected operators in the dual 1d CFT with arbitrary charges by lifting the lowest charge correlator to 4d. Note that the dual CFT is defined only formally since we are not considering gravity in the bulk. In order to realise 4d conformal symmetry, a crucial role is played by Casimirs of the superconformal group $SU(1,1|2)$. For example, we must act with these Casimirs on free theory correlators before lifting to 4d, and after unmixing correlators in the 2-derivative sector, the resulting anomalous dimensions turn out to be eigenvalues of the Casimirs. 

The second approach we develop is based on a 4d scalar effective action analogous to the 10d effective action for IIB string theory in AdS$_5 \times$S$^5$ \cite{Abl:2020dbx}. Using the effective action, one can compute generalised Witten diagrams which treat AdS and S on equal footing, giving a whole tower of 4-point correlators corresponding to harmonics on the sphere. Whereas the 10d effective action describes $\alpha'$ corrections to supergravity, the 4d effective action treats the 2-derivative sector (which arises from dimensional reduction of supergravity) on equal footing with higher derivative corrections. We find that the predictions of the 4d effective action are in perfect agreement with those of 4d conformal symmetry in the 2-derivative sector, but are generally incompatible with 4d conformal symmetry for higher derivative corrections except for a special infinite class of correlators. Moreover, after unmixing the correlators obtained from the higher-derivative effective action, we find that the resulting anomalous dimensions are no longer rational which further suggests that 4d conformal symmetry is broken. 

We then apply the lessons we learned in 1d to $\mathcal{N}=4$ SYM, for which there is a lot more data to compare to. Using superconformal Casimirs of $SU(2,2|4)$, we find that $\alpha'^3$ corrections enjoy 10d conformal symmetry, similarly to 2-derivative interactions in AdS$_2\times$S$^2$, while $\alpha'^5$ corrections generically break the 10d conformal symmetry, similarly to 6-derivative corrections in AdS$_2\times$S$^2$. So it is not in general possible to reproduce the predictions of the 10d effective action by acting with Casimirs on 10d conformal blocks beyond $\alpha'^3$,  except for a special class of correlators. We do on the other hand  find that there exists a 6-derivative action (corresponding to a four-derivative scalar effective action) for which the integrand of the corresponding  Witten diagrams has 10d conformal symmetry, although this combination does not appear to be consistent with the results of localisation in $\mathcal{N}=4$ SYM at $O(\alpha'^5)$.

There are a number of directions for future work. First of all it would be interesting to have a more systematic understanding of the (breaking of) 10d conformal symmetry by stringy corrections in AdS$_5 \times$S$^5$. Since $\alpha'^3$ corrections enjoy 10d conformal symmetry and are described by a $\phi^4$ interaction in the effective action, the breaking seems to be associated with derivative interactions in the effective action. It would be interesting to derive the appearance of the sixth order Casimir at $O(\alpha'^3)$ from first principles using a Weyl transformation from AdS$_5 \times$S$^5$ to flat space, and to use similar reasoning to understand the breaking of 10d conformal symmetry at $O(\alpha'^5)$ more systematically. In \cite{Caron-Huot:2018kta}, 10d conformal symmetry was also used to deduce the leading logarithmic part of all 4-point correlators of $1/2$-BPS operators in $\mathcal{N}=4$ SYM to all loops in the supergravity approximation, so it would be interesting to use 4d conformal symmetry to do the analogue of this for 1d correlators, which we expect to have a simpler structure. For example, whereas 1-loop correlators in $\mathcal{N}=4$ SYM have transcendentality four, recent results \cite{Ferrero:2019luz} suggest that they should have transcendentality two in 1d.  

Moreover since AdS$_2 \times$S$^2$ describes the near-horizon geometry of extremal 4d black holes, we hope this work will provide new tools for studying realistic black holes. As a first step, it would be interesting to study correlators of graviton multiplets AdS$_2 \times$S$^2$ and their relation to $\mathcal{N}=4$ SYK models \cite{Gates:2021iff,Gates:2021jdm}, $\mathcal{N}=4$ JT supergravity and the associated $\mathcal{N}=4$ Schwarzian theory deduced in \cite{Heydeman:2020hhw}. Another important aspect of AdS$_2$ is that it has two disconnected timelike boundaries. In this paper, we have only considered correlators on a single boundary, but it would be interesting to generalise our results to correlators connecting the two boundaries and explore the effects of higher-derivative corrections in this context, which may have implications for black hole entropy \cite{Azeyanagi:2007bj,Carroll:2009maa} or the weak gravity conjecture \cite{Arkani-Hamed:2006emk,Cheung:2018cwt}.

\begin{center}
\textbf{Acknowledgements}
\end{center}
We thank Francesco Aprile, Simon Caron-Huot, Shai Chester, Frank Coronado, James Drummond, Matthew Hydeman, Madalena Lemos, Hynek Paul and Michele Santagata for useful conversations. TA is supported by a Durham Doctoral Studentship, AL by the Royal Society as a Royal Society University Research Fellowship holder, and PH by an STFC Consolidated Grant ST/P000371/1. We also acknowledge support from  the European Union's Horizon 2020 research and innovation programme
under the Marie Sklodowska-Curie grant agreement No.764850 ``SAGEX''.

\appendix

\section{Casimirs}
\label{casimirderivation}

In this Appendix we will derive various properties of superconformal Casimir operators in 1d and 4d.

\subsection{The quadratic Casimir in $SU(1,1|2)$}

Superblocks are eigenfunctions of the quadratic super Casimir at points 1 and 2 acting on the correlator. To consider this we develop the formalism slightly in order to consider the coordinates $X_i$ explicitly  as sections on the super Grassmannian of $1|1$ planes inside a $2|2$ dimensional vector space. Note that the formalism and super Casimir outlined below generalises naturally from the supergroup $SU(1,1|2)$ to any supergroup of the form $SU(m,m|2n)$~\cite{Doobary:2015gia}.

Our space is the  Grassmannian  $Gr(1|1,2|2)$ , the space of $1|1\times 2|2$ matrices  $u^{\alpha}{}_{A}$. Here the small Greek indices refer to the local isotropy group $GL(1|1)$ whilst the big Latin indices refer to the global group $GL(2|2)$. Explicitly, one can put coordinates on this Grassmannian as
\begin{align}(u_i)^{\alpha}{}_{B}=\left(\delta^{\alpha}_{\beta},(X_i)^{\alpha}{}_{\dot{\beta}}\right)\text{ , }\bar{u}^{B}{}_{\dot{\alpha}}=\left(\begin{array}{c}-(X_i)^{\beta}{}_{\dot{\alpha}}\\ \delta^{\dot{\beta}}_{\dot{\alpha}}\end{array}\right),  \quad (X_i)^{\alpha}{}_{\dot{\beta}} = \left(	\begin{array}{cc}
x_i& \theta_i\\
\bar \theta_i & y_i
\end{array}
\right)\ .
\end{align}
We thus have $u^{\alpha}_{iB}\bar{u}^{B}_{j\dot{\alpha}}=X_{ij\dot{\alpha}}^{\alpha}$. Then the generators of the superconformal group, $SU(1,1|2)$, (at point $i$) are given simply as \begin{align}D^{A}_{iB}=u^{\alpha}_{iA}\frac{\partial}{\partial u^{\alpha}_{iB}}.\end{align}
The quadratic Casimir operator acting on the 4-point function at points $1$ and $2$ is then given as
\begin{align}\label{eq:cas}
{\mathcal C}_{1,2}^{SU(1,1|2)}=\frac12(D_{1B}^{A}+D_{2B}^{A})(D_{1A}^{B}+D_{2A}^{B}).\end{align}

Acting with the Casimir on the supercorrelator in the form~\eqref{corpre} and commuting through the prefactor gives
\begin{align}\label{eq:AA}&  {\mathcal C}_{1,2}^{SU(1,1|2)} \langle  \Psi_{p_1} \Psi_{p_2}  \Psi_{p_3} \Psi_{p_4}\rangle
=P_{p_i}\times\left[\left(\left(p_{12}{-}p_{34}\right)\left(x^2\frac{\partial}{\partial x}+y^2\frac{\partial}{\partial y}\right)+p_{34}p_{12}(x-y)\right)f+\frac12 {\mathcal C}_{1,2}^{SU(2|2)}f\right].\end{align}
To obtain this we used that $\text{sdet}(M)=\text{exp}(\text{str} \log(M))$ to deal with differentiating the propagators $g_{ij}:=\text{sdet}(X_{ij}^{-1})$ and then just applied the double derivative directly using  $D_{12B}^{A}u_{iC}^{\alpha}=u_{iB}^{\alpha}\delta^{A}_{C}\text{ and }D_{12A}^{B}\bar{u}_{i\dot{\delta}}^{C}=-\delta^{C}_{A}\bar{u}_{i\dot{\delta}}^{B}$ for $i=1$ or $2$.

We now consider the Casimir acting on $ f(Z)=f(x,y)=$ the conjugation invariant function of the cross-ratio matrix $Z$, which is thus a function of the eigenvalues of $Z$ only as discussed above~\eqref{strsdet}. By examining the action of the Casimir  on arbitrary products of traces of powers of $Z$,  $\prod_{i}\text{tr}(Z^i)^{a_{i}}$, and the corresponding expressions as polynomials of eigenvalues  we find that
\begin{align}
 {\mathcal C}_{1,2}^{SU(2|2)}f(x,y)&= \left[\left(x^2\frac{\partial}{\partial x}  -  \frac{2xy}{x-y} \right) (1-x)\frac{\partial}{\partial x} 
- \left(y^2\frac{\partial }{\partial y}  -  \frac{2yx}{y-x} \right) (1-y)\frac{\partial }{\partial y}\right]f(x,y)\\
&= \frac {x-y}{xy}\left(x^2 \partial_x(1-x)\partial_x -y^2 \partial_y(1-y)\partial_y\right)\frac {xyf(x,y)}{x-y}
\end{align}
Putting this together, we then obtain the action of the quadratic super Casimir on the correlator: 
\begin{align}\label{Delta2}
{\mathcal C}_{1,2}^{SU(1,1|2)}&= P_{p_i} \times \frac{x-y}{xy}   \Delta^{(2)}      \frac{xy}{x-y}     P_{p_i}^{-1}\ \notag \\
\Delta^{(2)} &= {\mathcal D}^{(p_{12},p_{43})}_x - {\mathcal D}^{(-p_{12},-p_{43})}_y \notag\\
{\mathcal D}^{(p_{12},p_{43})}_x &=x^2 \partial_x(1-x)\partial_x+(p_{12}+p_{43})x^2\partial_x - p_{12}p_{43} x\ .
\end{align}

Similar (simpler) computations give us the Casimirs of the subgroups $SU(1,1)$ and $SU(2)$ acting on the correlator as
\begin{align}
	{\mathcal C}_{1,2}^{SU(1,1)}&= P_{p_i} {\mathcal D}^{(p_{12},p_{43})}_x P_{p_i}^{-1}\notag\\
	{\mathcal C}_{1,2}^{SU(2)}&= P_{p_i} {\mathcal D}^{(-p_{12},-p_{43})}_y P_{p_i}^{-1},
\end{align} 
and  so we can write the action of  superconformal Casimir on the correlator directly in terms of the Casimir of the subgroups:
\begin{align}\label{casimirs}
{\mathcal C}_{1,2}^{SU(1,1|2)}&=   \frac{x-y}{xy}  \left( {\mathcal C}_{1,2}^{SU(1,1)} - {\mathcal C}_{1,2}^{SU(2)} \right)    \frac{xy}{x-y}\notag \\ &=   \cI  \left( {\mathcal C}_{1,2}^{SU(1,1)} - {\mathcal C}_{1,2}^{SU(2)} \right)    \cI^{-1}
\ ,
\end{align}
where in the second line we recall~\eqref{I} and use that $x_{12},x_{34},y_{12},y_{34}$ commute with ${\mathcal C}_{12}$. 

\subsection{Higher order Casimirs in  $SU(m,m|2n)$ CFTs}
\label{hocs}

The above formulae in fact generalise very nicely in two ways (see~\cite{fh} for more details). All {\em higher order} Casimirs for 4-point functions in an $SU(m,m|2n)$ CFT for {\em arbitrary} $m,n$ (where we are interested in $m=n=1$ for the 1d CFT case and $m=n=2$ for the $\cN=4$ SYM case) are given in terms of
{supersymmetric functions} of the differential operators $\cD_z^\pm$ defined in~\eqref{Delta2def}. 
A {symmetric function} $f(x_1,..,x_n)$ is one which is invariant under permutations of the $n$ variables. It can equivalently be viewed as a function of an $n \times n$ matrix $M$ invariant under conjugation, $f(M)=f(G^{-1}MG)$. The equivalence of these descriptions occurs  by identifying the variables $x_i$ with the eigenvalues of $M$. Similarly a { supersymmetric function} can be viewed as a function of an $(m|n)\times (m|n)$ supermatrix $M$ invariant under conjugation  $f(M)=f(G^{-1}MG)$. In terms of the eigenvalues of $M$ which we label $(x_1,..,x_m|y_1,..,y_n)$ the space of such functions is equivalent to the space of  doubly symmetric functions (symmetric separately in the two sets of variables $x_i$ and $y_j$) which also satisfies the additional constraint $\partial f (t,x_2,..|t,y_2,..)/\partial t=0$ (see e.g.~\cite{website} and references therein). A nice basis for the supersymmetric functions is provided by super Schur polynomials.

Any higher order  Casimir for a 4-point function in  an $SU(m,m|2n)$ theory (or more precisely its action on conformally invariant functions) is simply a supersymmetric function $f$ of appropriate differential operators $\cD_{z}^\pm$, conjugated with a  super Vandermonde determinant (and appropriate prefactor):
\begin{align}\label{hocs2}
	f({x_1},..,{x_m}|{y_1},..,{y_n}) &\rightarrow 	P_{p_i}V^{-1}f(\cD_{x_1}^+,..,\cD_{x_m}^+|\cD_{y_1}^-,..,\cD_{y_n}^-)VP_{p_i}^{-1},
\end{align}
where the super Vandermonde determinant is
\begin{align}
	V &= \frac{\prod_{1\leq i<j \leq m}(x_i^{-1}-x_j^{-1})\prod_{1\leq i<j \leq n}(y_i^{-1}-y_j^{-1})}{ \prod_{\substack{1\leq i\leq m\\ 1 \leq j\leq n}}(x_i^{-1}-y_j^{-1})}\ .
\end{align}
The bosonic case ($n=0$) of this result  follows from the results of Shimeno~\cite{shimeno} for $BC_m$ hypergeometric functions (see theorem 2.1)  following the identification  of the higher order Casimirs for blocks with the defining system of higher order differential equations for these hypergeometric functions in~\cite{fh}.
The supersymmetric generalisation is then naturally given by the above expression. 
The corresponding eigenvalues for superblocks of this Casimir are also given in~\cite{fh}.
 
\subsubsection*{Higher order Casimirs in the 1d SCFT}
The supersymmetric functions
in the case $m=n=1$ (relevant for the 1d SCFT) are straightforward to classify: they take the form
\begin{align}
	f^{(a,b)}(x|y) &= (x-y)x^a y^b\ ,
\end{align}
together with the trivial constant function.
Thus we classify all higher order Casimirs in the 1d SCFT as
\begin{align}\label{Deltaab}
	{\mathcal C}_{1,2}^{(a,b);SU(1,1|2)}&= \hat P_{p_i}  \frac{x-y}{xy} \,  \Delta^{(a,b)}   \,   \frac{xy}{x-y}     \hat P_{p_i}^{-1}\ \notag \\
	\Delta^{(a,b)} &= \left({\mathcal D}^+_x - {\mathcal D}^-_y \right)({\mathcal D}^+_x)^a({\mathcal D}^-_y)^b\ .
\end{align}
The case $a=b=0$ is then precisely the 1d quadratic Casimir of~\eqref{Delta2def} and higher order Casimirs in 1d are considered in~\eqref{fourthordercasimir} and the following.

Let us compare this with the higher order Casimirs for the maximal bosonic subgroup $SU(1,1)\times SU(2)$. These correspond to the product of the cases $(m,n)=(1,0)$ and $(m,n)=(0,1)$ in the general case~\eqref{hocs2} and take the form
\begin{align}
	{ C}_{1,2}^{(a,b);SU(1,1)\times SU(2)}&= \hat P_{p_i} \, ({\mathcal D}^+_x)^a({\mathcal D}^-_y)^b  \, \hat P_{p_i}^{-1}.\ 
\end{align}
We then find the relation
\begin{align}
	{\mathcal C}_{1,2}^{(a,b)}= \frac{x-y}{xy}  \left(	{ C}_{1,2}^{(a+1,b)}- 	{ C}_{1,2}^{(a,b+1)}\right) \frac{xy}{x-y} \ ,
\end{align}
of which~\eqref{casimirs} is a special case. 

 \subsubsection*{Higher order Casimirs in $\cN=4$ SYM}
 
The space of supersymmetric functions in the case $m=n=2$ (relevant for $\cN=4$ SYM) is more intricate, but a nice basis for them is given by the super Schur polynomials, labelled by Young diagrams. They fall into two classes: short and long. We only discuss the long ones as they are the only ones needed here, but all cases, both long and short, can be found in~\cite{Heslop:2002hp}. The long supersymmetric functions in the case $m=n=2$ are spanned  by the  functions%
\footnote{This basis is equivalent to but not identical to  the Schur polynomial  basis for long supersymmetric polynomials.}
\begin{align}
	f^{(a,\bar a|b,\bar b)}(x,\bar x|y,\bar y) &= (x-y)(x-\bar y)(\bar x-y)(\bar x-\bar y) ( x^a \bar x^{\bar a} + x^{\bar a} \bar x^{ a}  ) ( y^b \bar y^{\bar b} + y^{\bar b} \bar y^{ b}  )\ .
\end{align}
Noting that the first four factors  translate directly to $\Delta^{(8)}$~\eqref{Delta8def} under the prescription~\eqref{hocs2}, we see that the  space of corresponding higher order Casimirs is spanned by the operators
\begin{align}\label{Deltaababbb}	
	{\mathcal C}_{1,2}^{(a,\bar a|b,\bar b);SU(2,2|4)}&=  \hat P_{p_i}\left(\frac{(x{-}y)(x{-}\bar y)(\bar x{-}y)(\bar x{-}\bar y)}{x \bar x y \bar y (x{-}\bar x)(y {-}\bar y)}\right)    \Delta^{(8)}\Delta^{(a,\bar a|b,\bar b)}    \left(\frac{x \bar x y \bar y (x{-}\bar x)(y {-}\bar y)}{(x{-}y)(x{-}\bar y)(\bar x{-}y)(\bar x{-}\bar y)}\right)    \hat P_{p_i}^{{-}1}\\
\Delta^{(a,\bar a|b,\bar b)} &=  \Big( (\cD_{ x}^+)^a (\cD_{\bar x}^+)^{\bar a} + (\cD_{ x}^+)^{\bar a} (\cD_{\bar x}^+)^{ a}  \Big) \Big( (\cD_{ y}^-)^b (\cD_{\bar y}^-)^{\bar b} + (\cD_{ y}^-)^{\bar b} (\cD_{\bar y}^-)^{ b}  \Big)\label{Deltaaabbbb}\ .
\end{align}
This is the basis of higher order Casimirs used to find the 14th order Casimir in the $\alpha'^3$ correction~\eqref{preamp10d}-\eqref{delta6} as well as the higher order Casimirs considered at higher orders in $\alpha'$ in section~\ref{alpha5}.

Let us compare this super Casimir with the higher order Casimirs for the maximal bosonic subgroup $SU(2,2)\times SU(4)$ which are obtained from the general formula~\eqref{hocs2} by taking the product of the cases $(m,n)=(2,0)$ and $(m,n)=(0,2)$ and take the form
\begin{align}\label{nis4boshocs}
	{ C}_{1,2}^{(a,\bar a |b,\bar b);SU(2,2)\times SU(4)}&= \hat P_{p_i} \left(\frac{x \bar x y \bar y}{ (x{-}\bar x)(y {-}\bar y)}\right)  \, \Delta^{(a,\bar a|b,\bar b)} \, \left(\frac{(x{-}\bar x)(y {-}\bar y)}{x \bar x y \bar y }\right) \,\hat P_{p_i}^{{-}1},\ 
\end{align}
where $\Delta^{(a,\bar a|b,\bar b)}$ is in~\eqref{Deltaaabbbb}.
We find that super Casimirs are related to the bosonic Casimirs via
\begin{align}\label{superctoboscnis4}
{\mathcal C}_{1,2}^{(a,\bar a|b,\bar b);SU(2,2|4)}= \left(\frac{(x{-}y)(x{-}\bar y)(\bar x{-}y)(\bar x{-}\bar y)}{(x \bar x y \bar y)^2}\right)	{ C}_{1,2}^{(*);SU(2,2)\times SU(4)}  \left(\frac{(x \bar x y \bar y)^2}{(x{-}y)(x{-}\bar y)(\bar x{-}y)(\bar x{-}\bar y)}\right),
\end{align}
 where ${ C}_{1,2}^{(*)}$ is a linear combination of the bosonic Casimirs in~\eqref{nis4boshocs} obtained by multiplying by $\Delta^{(8)}$ and expanding, so the first and last terms for example are 
 \begin{align}
 	{ C}_{1,2}^{(*)} = 	{ C}_{1,2}^{(a+2,\bar a+2 |b,\bar b)} - \dots  + { C}_{1,2}^{(a,\bar a |b+2,\bar b+2)}\ .
 \end{align}

\section{Correlators of descendants}
\label{sec:desc}

In this Appendix we wish to compute the correlator of superconformal descendants (see \eqref{Psi}): 
\begin{align}
\partial_{\theta_1}\partial_{\theta_2}\partial_{\bar \theta_3}\partial_{\bar \theta_4} \langle  \Psi_{p_1} \Psi_{p_2}  \Psi_{p_3} \Psi_{p_4}\rangle|_{\theta_i=\bar \theta_i=0}
\end{align}
 and prove the result~\eqref{descendentcorrelator1}.
 
 We must first understand the general form the supercorrelator must take according to the superconformal Ward identities.
Superconformal symmetry $SU(1,1|2)$ fixes the correlator in terms of a conjugation invariant function of a cross ratio matrix (see~\cite{Doobary:2015gia,Heslop:2002hp})
The general arguments in~\cite{Doobary:2015gia,Heslop:2002hp} show that superconformal symmetry $SU(1,1|2)$ fixes the correlator to $P_{p_i} \times f(Z)$ where  $f(Z)$ is a conjugation invariant function ($f(Z)=f(G^{-1}ZG)$) of the cross-ratio matrix $Z$
\begin{align}
	Z:=X_{12}X_{24}^{-1}X_{43}X_{31}^{-1}=\left(	\begin{array}{cc}
		x& \xi\\
		\bar \xi & y
	\end{array}
	\right),\qquad X_{ij}:=X_i-X_j, \qquad 
	X_i:=\left(	\begin{array}{cc}
		x_i& \theta_i\\
		\bar \theta_i & y_i
	\end{array}
	\right)\ .
\end{align}
It thus becomes a function of the eigenvalues of $Z$ only, $\hat x, \hat y$
so
\begin{align}\label{corpre}
	\langle  \Psi_{p_1} \Psi_{p_2}  \Psi_{p_3} \Psi_{p_4}\rangle= P_{p_i} \times f(\hat x, \hat y)
\end{align}
where
\begin{align}\label{zhwh}
\hat x = x +\frac{\xi \bar \xi}{x-y} \qquad 	\hat y = y +\frac{\xi \bar \xi}{x-y}
\end{align}
That the eigenvalues of $Z$ are given by \eqref{zhwh} can be easily checked by verifying that they give the same super trace and super determinant: 
\begin{align}\label{strsdet}
	\text{str}(Z)=x-y=\hat x-\hat y, \qquad  \text{sdet}(Z)=(x-\xi\bar \xi/y)/y= \hat x/\hat y.
\end{align}

Plugging these definitions into Mathematica with the help of the \texttt{Grassmann} package~\cite{grassmann} to deal with the Grassmann odd variables one finds that the Grassmann odd derivatives acting on $f$ are consistent with the following differential operator:
\begin{align}
	\partial_{\theta_1}\partial_{\theta_2}\partial_{\bar \theta_3}\partial_{\bar \theta_4} f(\hat x,\hat y)|_{\theta_i=\bar \theta_i=0} = \frac1{x_{12}x_{34}y_{12}y_{34}} \left({\mathcal D}^{(0,0)}_{\hat x} - {\mathcal D}^{(0,0)}_{\hat y} \right) \frac{\hat x \hat y}{\hat x-\hat y}f(\hat x,\hat y)\ .
\end{align}
Pulling this through the prefactor then gives the action on the correlator itself: 
\begin{align}
&\partial_{\theta_1}\partial_{\theta_2}\partial_{\bar \theta_3}\partial_{\bar \theta_4} \langle  \Psi_{p_1} \Psi_{p_2}  \Psi_{p_3} \Psi_{p_4}\rangle|_{\theta_i=\bar \theta_i=0} \notag \\&= \frac {P_{p_i}}{x_{12}x_{34}y_{12}y_{34}} \left({\mathcal D}^{(p_{12},p_{43})}_{ x} - {\mathcal D}^{(-p_{12},-p_{43})}_{ y} \right) P_{p_i}^{-1} \frac{x y}{ x- y}\langle  \Psi_{p_1} \Psi_{p_2}  \Psi_{p_3} \Psi_{p_4}\rangle|_{\theta_i=\bar \theta_i=0}\notag \\
&= {\mathcal I}^{-1} {\mathcal C}_{1,2}^{SU(1,1|2)}\langle  \Psi_{p_1} \Psi_{p_2}  \Psi_{p_3} \Psi_{p_4}\rangle|_{\theta_i=\bar \theta_i=0}
\ ,
\end{align}
with $\mathcal I$ defined in~\eqref{I}. 
This is precisely the result claimed in~\eqref{descendentcorrelator1}.  It would be interesting to understand in a more direct way precisely why the Casimir appears in this context. 

\section{Further Unmixing Examples} \label{unmixingexamples}

In the end of section \ref{sec:unmixing} we described some simple examples of unmixing in the 2-derivative sector. In this Appendix, we will describe more complicated examples. First we will describe unmixing in the 2-derivative sector with non-zero $SU(2)$ charge $p$ and then we will describe some examples in the 6-derivative sector.

\subsection{2-derivative sector}

The simplest case with $p>0$ is $\left(\Delta,p\right)=(2,1)$, for which there is only one exchanged operator in the double-trace spectrum, $\cO_1\cO_2$. Performing the conformal block expansion of $H_{1212}$ and comparing to \eqref{conformalblockexpansion1/c}, we find that the coefficients for $\left(\Delta,p\right)=(2,1)$ are
\begin{align}\label{freetheorysugra21}
A^{(0)}_{1212}(2,1)&=\frac{1}{4}\ ,\quad M^\text{(2)}_{1212}(2,1)=1.
\end{align}
The unmixing equations in \eqref{unmixingequations} are then solved by 
\begin{equation}\label{unmixedcoeffssugra21}
\gamma^\text{(2)}_{2,1}=4\ ,\quad C^{(0)}_{2,1}=\frac{1}{2}\ .
\end{equation}

Next, let us solve the mixing problem for $(\Delta,p)=(4,1)$ where two types of operators are exchanged, $\cO_{1}\partial \cO_{2}$ and $\cO_{2} \cO_{3}$. Performing the conformal block expansion of the relevant correlators then gives
\begin{align}\label{freetheorysugra41}
\hat{A}^{(0)}_{4,1}=
\begin{pmatrix}
A^{(0)}_{1212} & 0 \\
0 & A^{(0)}_{2323}
\end{pmatrix}_{(4,1)}=\begin{pmatrix}
\frac{5}{84}  & 0 \\
 0  & \frac{1}{30} 
\end{pmatrix}\ ,\ \hat{M}^\text{(2)}_{4,1}=
\begin{pmatrix}
M^\text{(2)}_{1212} & M^\text{(2)}_{1223} \\
M^\text{(2)}_{1223} & M^\text{(2)}_{2323}
\end{pmatrix}_{(4,1)}=\begin{pmatrix}
\frac{5}{21}  & \frac{1}{3} \\
 \frac{1}{3}  & \frac{7}{15} 
\end{pmatrix}\ .
\end{align}
Solving the unmixing equations gives the following anomalous dimensions and OPE coefficients:
\begin{align}\label{unmixedcoeffssugra41}
\gamma^\text{(2)}_{4,1}=18\ ,\quad \mathbb{C}^{(0)}_{4,1}=\begin{pmatrix}
\frac{\sqrt{5}}{3\sqrt{42}} & -\frac{\sqrt{5}}{6\sqrt{3}} \\
\frac{\sqrt{7}}{3\sqrt{30}} & \frac{1}{3\sqrt{15}}
\end{pmatrix}\ .
\end{align}
Note that there is only one non-zero anomalous dimension, as expected.

\subsection{6-derivative sector}

Let us first consider $t=\Delta-p$ odd. For the simplest case $\left(\Delta,p\right)=(1,0)$ only one exchanged operator $\cO_{1} \cO_{1}$ contributes. Performing the conformal block expansion of $H^\text{(6)}_{1111}$ in \eqref{fourdercorrelators}, we find that the coefficient for $\left(\Delta,p\right)=(1,0)$ is 
\begin{align}
M^\text{(6)}_{1111}(1,0)=4\ .
\end{align}
The unmixing equations are
\begin{equation}
 M^\text{(6)}_{1111}(1,0)=\gamma^\text{(6)}_{1,0}\left(C^{(0)}_{1,0}\right)^2+2\,\gamma^\text{(2)}_{1,0}C^{(0)}_{1,0}C^\text{(6)}_{1,0}\ ,\quad 0=C^{(0)}_{1,0}C^\text{(6)}_{1,0}\\ .
\end{equation}
We can solve these equations using \eqref{unmixedcoeffssugra10}, which yields
\begin{equation}
\left(\gamma^\text{(6)}_A\right)^{1,0}_{0,0}=4\ ,\quad C^\text{(6)}_{1,0}=0\ .
\end{equation}

For $\left(\Delta,p\right)=(3,0)$ there are two possible exchanged operators, $\cO_{2}\cO_{2}$ and $\cO_{1} \partial^2 \cO_{1}$. In this case we find
  \begin{align}
\hat{M}^\text{(6)}_{3,0}=
\begin{pmatrix}
M^\text{4-deriv}_{1111} & M^\text{4-deriv}_{1122} \\
M^\text{4-deriv}_{1122} & M^\text{4-deriv}_{2222}
\end{pmatrix}_{(3,0)}=\begin{pmatrix}
\frac{72}{5} & \frac{4}{3}(15+\,C_0) \\
 \frac{4}{3}(15+\,C_0) & \frac{8}{9}(38+5\,C_0)
\end{pmatrix}.\ 
\end{align}
Solving the unmixing equations using ~\eqref{unmixedcoeffssugra30} then gives
\begin{align}
\gamma^\text{(6)}_{3,0,i}&=\left\{\frac{16}{3}(137+15 C_0 ),\ \frac{64}{3}\right\}\ ,\ \mathbb{C}^\text{(6)}_{3,0}=
(-1+3 C_0)\begin{pmatrix}
\frac{-\sqrt {5}}{9\sqrt{3}} & \frac{1}{9\sqrt{3}} \\
\frac{\sqrt{5}}{27 \sqrt{3}} & \frac{5}{27 \sqrt{3}}
\end{pmatrix}\ .
\end{align}
Recall that $C_0$ is the coefficient of the ambiguity in~\eqref{fourderivSeff}. While there is only one non-zero anomalous dimension for each $\left(\Delta,p\right)$ in the 2-derivative sector, we see that there are two non-zero anomalous dimensions in the 6-derivative sector for $(\Delta,p)=(3,0)$.

The first example where square roots appear in the 6-derivative sector is $(\Delta,p)=(5,2)$. In this case there are four class $A$ operators, $\left\{\cO_1\partial^2\cO_3,\cO_2\cO_4,\cO_2\partial^2\cO_2,\cO_3\cO_3 \right\}$, and one class $B$ operator, $\cO_2\partial\cO_3$. Class $A$ and $B$ operators do not mix with each other and the class $A$ operators will have square roots in their anomalous dimensions after unmixing. Performing the conformal block expansion of the relevant correlators at $\cO(a/c)$ then leads to the symmetric matrix:
{\small
\begin{align}\label{fourderiv52matrices}
&\hat{M}^\text{(6)}_{5,2}=
\begin{pmatrix}
M^\text{(6)}_{1313} & M^\text{(6)}_{1324} & M^\text{(6)}_{1322} & M^\text{(6)}_{1333} & M^\text{(6)}_{1323} \\
 & M^\text{(6)}_{2424} & M^\text{(6)}_{2422} & M^\text{(6)}_{2433} & M^\text{(6)}_{2423} \\
 &  & M^\text{(6)}_{2222} & M^\text{(6)}_{2233} & M^\text{(6)}_{2223} \\
 &  &  & M^\text{(6)}_{3333} & M^\text{(6)}_{3323} \\
 &  &  &  & M^\text{(6)}_{2323} 
\end{pmatrix}_{(5,2)}\notag\\
&=\begin{pmatrix}
\frac{2}{25}(774+35 C_0)  & \frac{6}{35}(416+35 C_0) & \frac{4}{45}(942+35 C_0) & \frac{36}{35}(90+7 C_0) & 0 \\
   & \frac{6}{245}(3910+441 C_0) &  \frac{36}{35}(90+7 C_0) & \frac{468}{245}(62+7 C_0) & 0\\
 &  & \frac{8}{135}(1955+56 C_0) & \frac{8}{105}(1707+112 C_0) & 0 \\
  & & & \frac{72}{245}(657+56 C_0) & 0\\
  & & & & \frac{672}{25}
\end{pmatrix}\ .
\end{align}
}
Solving the unmixing equations then gives
\begin{align}
\left(\gamma^\text{(6)}_A\right)^{5,2}_{0,0}=&\frac{32}{3}(545+51 C_0)\ ,\quad \left(\gamma^\text{(6)}_A\right)^{5,2}_{1,0}=\frac{8}{15}(415-\sqrt{2881})\ ,\notag\\ \left(\gamma^\text{(6)}_A\right)^{5,2}_{0,1}=&\frac{8}{15}(415+\sqrt{2881})\ ,\quad \left(\gamma^\text{(6)}_{B^\text{to}}\right)^{5,2}_1=\frac{1008}{5}\ .
\end{align}
Note the appearance of square roots, which signal the breaking of 4d conformal symmetry.

Finally, let us describe a few examples of unmixing for $t=\Delta-p$ even. At weight $(\Delta,p)=(3,1)$ there is only one operator present, $\cO_1\partial\cO_2$, and we get
\begin{align}
A^{(0)}_{1212}(3,1)&=\frac{2}{15}\ ,\ M^\text{(6)}_{1212}(3,1)=\frac{16}{9}\ .
\end{align}
Solving the unmixing equations then gives the anomalous dimension and OPE coefficients
\begin{equation}
\left(\gamma^\text{(6)}_{B^\text{te}}\right)^{3,1}_{1,0}=\frac{40}{3}\ ,\quad C^{(0)}_{3,1}=\sqrt{\frac{2}{15}}\ ,\quad C^{\text{(6)}}_{3,1}=0\ .
\end{equation}
For $(\Delta,p)=(5,1)$ there are two operators that contribute, $\cO_1\partial^2\cO_2$ and $\cO_2\partial\cO_3$, and performing the conformal block expansion of the relevant correlators gives
\begin{align}
\hat{A}^{(0)}_{5,1}&=\begin{pmatrix}
A^{(0)}_{1212} & 0\\
0 & A^{(0)}_{2323}\\
\end{pmatrix}_{(5,1)}=\begin{pmatrix}
\frac{1}{42} & 0\\
0 & \frac{9}{175}
\end{pmatrix}\ ,\notag\\
\hat{M}^\text{(6)}_{5,1}&=\begin{pmatrix}
M^\text{(6)}_{1212} & M^\text{(6)}_{1223}\\
M^\text{(6)}_{1223} & M^\text{(6)}_{2323}\\
\end{pmatrix}_{(5,1)}=\begin{pmatrix}
\frac{52}{45} & \frac{104}{25}\\
\frac{104}{25} & \frac{1872}{125}
\end{pmatrix}\ .
\end{align}
Solving the unmixing equations for this case gives the anomalous dimension
\begin{equation}
\left(\gamma^\text{(6)}_{B^\text{te}}\right)^{5,1}_{1,0}=\frac{5096}{15}\ .
\end{equation}
For this case again, as for the whole $p=1$ sector, there is exactly one non-zero anomalous dimension, which corresponds to the operator with $i_{B^\text{te}}=1,\ r_{B^\text{te}}=0$. Going to $p\geq 2$, there will be two non-zero anomalous dimensions, the additional one being for the operator with $i_{B^\text{te}}=1,\ r_{B^\text{te}}=1$.

\raggedright


\begin{thebibliography}{99}
	
	\bibitem{Maldacena:1997re}
	J.~M.~Maldacena,
	``The Large N limit of superconformal field theories and supergravity,''
	Adv. Theor. Math. Phys. \textbf{2} (1998), 231-252
	doi:10.1023/A:1026654312961
	[arXiv:hep-th/9711200 [hep-th]].
	
	\bibitem{Maldacena:2002vr}
	J.~M.~Maldacena,
	``Non-Gaussian features of primordial fluctuations in single field inflationary models,''
	JHEP \textbf{05} (2003), 013
	doi:10.1088/1126-6708/2003/05/013
	[arXiv:astro-ph/0210603 [astro-ph]].
	
	\bibitem{Bertotti:1959pf}
	B.~Bertotti,
	``Uniform electromagnetic field in the theory of general relativity,''
	Phys. Rev. \textbf{116} (1959), 1331
	doi:10.1103/PhysRev.116.1331
	
	\bibitem{Robinson:1959ev}
	I.~Robinson,
	``A Solution of the Maxwell-Einstein Equations,''
	Bull. Acad. Pol. Sci. Ser. Sci. Math. Astron. Phys. \textbf{7} (1959), 351-352
	
	\bibitem{Caron-Huot:2018kta}
	S.~Caron-Huot and A.~K.~Trinh,
	``All tree-level correlators in AdS$_{5}$\texttimes{}S$_{5}$ supergravity: hidden ten-dimensional conformal symmetry,''
	JHEP \textbf{01} (2019), 196
	doi:10.1007/JHEP01(2019)196
	[arXiv:1809.09173 [hep-th]].

      \bibitem{Rastelli:2017udc}
      L.~Rastelli and X.~Zhou,
     ``How to Succeed at Holographic Correlators Without Really Trying,''
     JHEP \textbf{04} (2018), 014
     doi:10.1007/JHEP04(2018)014
     [arXiv:1710.05923 [hep-th]].
	
	\bibitem{Rastelli:2019gtj}
	L.~Rastelli, K.~Roumpedakis and X.~Zhou,
	``$\mathbf{AdS_3\times S^3}$ Tree-Level Correlators: Hidden Six-Dimensional Conformal Symmetry,''
	JHEP \textbf{10} (2019), 140
	doi:10.1007/JHEP10(2019)140
	[arXiv:1905.11983 [hep-th]].
	
	\bibitem{Giusto:2020neo}
	S.~Giusto, R.~Russo, A.~Tyukov and C.~Wen,
	``The CFT$_6$ origin of all tree-level 4-point correlators in AdS$_3 \times S^3$,''
	Eur. Phys. J. C \textbf{80} (2020) no.8, 736
	doi:10.1140/epjc/s10052-020-8300-4
	[arXiv:2005.08560 [hep-th]].
	
	\bibitem{Aprile:2021mvq}
	F.~Aprile and M.~Santagata,
	``Two-particle spectrum of tensor multiplets coupled to $AdS_3\times S^3$ gravity,''
	[arXiv:2104.00036 [hep-th]].
	
	\bibitem{Wen:2021lio}
	C.~Wen and S.~Q.~Zhang,
	``Notes on gravity multiplet correlators in $AdS_3 \times S^3$,''
	JHEP \textbf{07} (2021), 125
	doi:10.1007/JHEP07(2021)125
	[arXiv:2106.03499 [hep-th]].

      \bibitem{Alday:2021odx}
      L.~F.~Alday, C.~Behan, P.~Ferrero and X.~Zhou,
      ``Gluon Scattering in AdS from CFT,''
      JHEP \textbf{06} (2021), 020
      doi:10.1007/JHEP06(2021)020
      [arXiv:2103.15830 [hep-th]].
	
	\bibitem{Drummond:2020dwr}
	J.~M.~Drummond, H.~Paul and M.~Santagata,
	``Bootstrapping string theory on AdS$_5 \times S^5$,''
	[arXiv:2004.07282 [hep-th]].
	
	\bibitem{Abl:2020dbx}
	T.~Abl, P.~Heslop and A.~E.~Lipstein,
	``Towards the Virasoro-Shapiro amplitude in AdS$_{5} \times S^{5}$,''
	JHEP \textbf{04} (2021), 237
	doi:10.1007/JHEP04(2021)237
	[arXiv:2012.12091 [hep-th]].
	
	\bibitem{Goncalves:2014ffa}
	V.~Gon\c{c}alves,
	``Four point function of $\mathcal{N}=4$ stress-tensor multiplet at strong coupling,''
	JHEP \textbf{04} (2015), 150
	doi:10.1007/JHEP04(2015)150
	[arXiv:1411.1675 [hep-th]].
	
	\bibitem{Alday:2018pdi}
	L.~F.~Alday, A.~Bissi and E.~Perlmutter,
	``Genus-One String Amplitudes from Conformal Field Theory,''
	JHEP \textbf{06} (2019), 010
	doi:10.1007/JHEP06(2019)010
	[arXiv:1809.10670 [hep-th]].
	
	\bibitem{Drummond:2019odu}
	J.~M.~Drummond, D.~Nandan, H.~Paul and K.~S.~Rigatos,
	``String corrections to AdS amplitudes and the double-trace spectrum of $ \mathcal{N} $ = 4 SYM,''
	JHEP \textbf{12} (2019), 173
	doi:10.1007/JHEP12(2019)173
	[arXiv:1907.00992 [hep-th]].
	
	\bibitem{Aprile:2020luw}
	F.~Aprile and P.~Vieira,
	``Large $p$ explorations. From SUGRA to big STRINGS in Mellin space,''
	JHEP \textbf{12} (2020), 206
	doi:10.1007/JHEP12(2020)206
	[arXiv:2007.09176 [hep-th]].
	
	\bibitem{Qiao:2017xif}
	J.~Qiao and S.~Rychkov,
	``A tauberian theorem for the conformal bootstrap,''
	JHEP \textbf{12} (2017), 119
	doi:10.1007/JHEP12(2017)119
	[arXiv:1709.00008 [hep-th]].

      \bibitem{Maldacena:1998uz}
      J.~M.~Maldacena, J.~Michelson and A.~Strominger,
      JHEP \textbf{02} (1999), 011
      doi:10.1088/1126-6708/1999/02/011
      [arXiv:hep-th/9812073 [hep-th]].
	
	\bibitem{Almheiri:2014cka}
	A.~Almheiri and J.~Polchinski,
	``Models of AdS$_{2}$ backreaction and holography,''
	JHEP \textbf{11} (2015), 014
	doi:10.1007/JHEP11(2015)014
	[arXiv:1402.6334 [hep-th]].
	
	\bibitem{Maldacena:2016upp}
	J.~Maldacena, D.~Stanford and Z.~Yang,
	``Conformal symmetry and its breaking in two dimensional Nearly Anti-de-Sitter space,''
	PTEP \textbf{2016} (2016) no.12, 12C104
	doi:10.1093/ptep/ptw124
	[arXiv:1606.01857 [hep-th]].
	
	\bibitem{Michelson:1999kn}
	J.~Michelson and M.~Spradlin,
	``Supergravity spectrum on AdS(2) x S**2,''
	JHEP \textbf{09} (1999), 029
	doi:10.1088/1126-6708/1999/09/029
	[arXiv:hep-th/9906056 [hep-th]].
	
	\bibitem{Lee:1999yu}
	J.~Lee and S.~Lee,
	``Mass spectrum of D = 11 supergravity on AdS(2) x S**2 x T**7,''
	Nucl. Phys. B \textbf{563} (1999), 125-149
	doi:10.1016/S0550-3213(99)00598-2
	[arXiv:hep-th/9906105 [hep-th]].
	
	\bibitem{Aprile:2017xsp}
	F.~Aprile, J.~M.~Drummond, P.~Heslop and H.~Paul,
	``Unmixing Supergravity,''
	JHEP \textbf{02} (2018), 133
	doi:10.1007/JHEP02(2018)133
	[arXiv:1706.08456 [hep-th]].
	
	\bibitem{Binder:2019jwn}
	D.~J.~Binder, S.~M.~Chester, S.~S.~Pufu and Y.~Wang,
	``$ \mathcal{N} $ = 4 Super-Yang-Mills correlators at strong coupling from string theory and localization,''
	JHEP \textbf{12} (2019), 119
	doi:10.1007/JHEP12(2019)119
	[arXiv:1902.06263 [hep-th]].
	
	\bibitem{Azeyanagi:2007bj}
	T.~Azeyanagi, T.~Nishioka and T.~Takayanagi,
	``Near Extremal Black Hole Entropy as Entanglement Entropy via AdS(2)/CFT(1),''
	Phys. Rev. D \textbf{77} (2008), 064005
	doi:10.1103/PhysRevD.77.064005
	[arXiv:0710.2956 [hep-th]].
	
	\bibitem{Doobary:2015gia}
	R.~Doobary and P.~Heslop,
	``Superconformal partial waves in Grassmannian field theories,''
	JHEP \textbf{12} (2015), 159
	doi:10.1007/JHEP12(2015)159
	[arXiv:1508.03611 [hep-th]].
	
	\bibitem{Galperin:1984av}
	A.~Galperin, E.~Ivanov, S.~Kalitsyn, V.~Ogievetsky and E.~Sokatchev,
	``Unconstrained N=2 Matter, Yang-Mills and Supergravity Theories in Harmonic Superspace,''
	Class. Quant. Grav. \textbf{1} (1984), 469-498
	[erratum: Class. Quant. Grav. \textbf{2} (1985), 127]
	doi:10.1088/0264-9381/1/5/004
	
	\bibitem{Howe:1995md}
	P.~S.~Howe and G.~G.~Hartwell,
	``A Superspace survey,''
	Class. Quant. Grav. \textbf{12} (1995), 1823-1880
	doi:10.1088/0264-9381/12/8/005
	
	\bibitem{Arutyunov:2002fh}
	G.~Arutyunov, F.~A.~Dolan, H.~Osborn and E.~Sokatchev,
	``Correlation functions and massive Kaluza-Klein modes in the AdS / CFT correspondence,''
	Nucl. Phys. B \textbf{665} (2003), 273-324
	doi:10.1016/S0550-3213(03)00448-6
	[arXiv:hep-th/0212116 [hep-th]].
	
	\bibitem{Mazac:2018qmi}
	D.~Maz\'a\v{c},
	``A Crossing-Symmetric OPE Inversion Formula,''
	JHEP \textbf{06} (2019), 082
	doi:10.1007/JHEP06(2019)082
	[arXiv:1812.02254 [hep-th]].
	
	\bibitem{Giombi:2017cqn}
	S.~Giombi, R.~Roiban and A.~A.~Tseytlin,
	``Half-BPS Wilson loop and AdS$_2$/CFT$_1$,''
	Nucl. Phys. B \textbf{922} (2017), 499-527
	doi:10.1016/j.nuclphysb.2017.07.004
	[arXiv:1706.00756 [hep-th]].
	
	\bibitem{Dolan:2000ut}
	F.~A.~Dolan and H.~Osborn,
	``Conformal four point functions and the operator product expansion,''
	Nucl. Phys. B \textbf{599} (2001), 459-496
	doi:10.1016/S0550-3213(01)00013-X
	[arXiv:hep-th/0011040 [hep-th]].
	
	\bibitem{Heemskerk:2009pn}
	I.~Heemskerk, J.~Penedones, J.~Polchinski and J.~Sully,
	``Holography from Conformal Field Theory,''
	JHEP \textbf{10} (2009), 079
	doi:10.1088/1126-6708/2009/10/079
	[arXiv:0907.0151 [hep-th]].
	
	\bibitem{Ferrero:2019luz}
	P.~Ferrero, K.~Ghosh, A.~Sinha and A.~Zahed,
	``Crossing symmetry, transcendentality and the Regge behaviour of 1d CFTs,''
	JHEP \textbf{07} (2020), 170
	doi:10.1007/JHEP07(2020)170
	[arXiv:1911.12388 [hep-th]].
	
	\bibitem{Aprile:2018efk}
	F.~Aprile, J.~Drummond, P.~Heslop and H.~Paul,
	``Double-trace spectrum of $N=4$ supersymmetric Yang-Mills theory at strong coupling,''
	Phys. Rev. D \textbf{98} (2018) no.12, 126008
	doi:10.1103/PhysRevD.98.126008
	[arXiv:1802.06889 [hep-th]].
	
	\bibitem{Aprile:2020mus}
	F.~Aprile, J.~M.~Drummond, H.~Paul and M.~Santagata,
	``The Virasoro-Shapiro amplitude in AdS$_{5}$ \texttimes{} S$^{5}$ and level splitting of 10d conformal symmetry,''
	JHEP \textbf{11} (2021), 109
	doi:10.1007/JHEP11(2021)109
	[arXiv:2012.12092 [hep-th]].
	
	\bibitem{Caron-Huot:2021usw}
	S.~Caron-Huot and F.~Coronado,
	``Ten dimensional symmetry of $N$ = 4 SYM correlators,''
	[arXiv:2106.03892 [hep-th]].
	
	\bibitem{Drummond:2006by}
	J.~M.~Drummond, L.~Gallot and E.~Sokatchev,
	``Superconformal Invariants or How to Relate 4-point AdS Amplitudes,''
	Phys. Lett. B \textbf{645} (2007), 95-100
	doi:10.1016/j.physletb.2006.12.015
	[arXiv:hep-th/0610280 [hep-th]].
	
	\bibitem{Chicherin:2015edu}
	D.~Chicherin, J.~Drummond, P.~Heslop and E.~Sokatchev,
	``All three-loop 4-point correlators of half-BPS operators in planar $ \mathcal{N} $ = 4 SYM,''
	JHEP \textbf{08} (2016), 053
	doi:10.1007/JHEP08(2016)053
	[arXiv:1512.02926 [hep-th]].
	
	\bibitem{Gates:2021iff}
	S.~J.~Gates, Y.~Hu and S.~N.~H.~Mak,
	``On 1D, N = 4 Supersymmetric SYK-Type Models (II),''
	[arXiv:2110.15562 [hep-th]].
	
	\bibitem{Gates:2021jdm}
	S.~J.~Gates, Y.~Hu and S.~N.~H.~Mak,
	``On 1D, N = 4 Supersymmetric SYK-Type Models (I),''
	doi:10.1007/JHEP06(2021)158
	[arXiv:2103.11899 [hep-th]].
	
	\bibitem{Heydeman:2020hhw}
	M.~Heydeman, L.~V.~Iliesiu, G.~J.~Turiaci and W.~Zhao,
	``The statistical mechanics of near-BPS black holes,''
	[arXiv:2011.01953 [hep-th]].
	
	\bibitem{Carroll:2009maa}
	S.~M.~Carroll, M.~C.~Johnson and L.~Randall,
	``Extremal limits and black hole entropy,''
	JHEP \textbf{11} (2009), 109
	doi:10.1088/1126-6708/2009/11/109
	[arXiv:0901.0931 [hep-th]].
	
	\bibitem{Arkani-Hamed:2006emk}
	N.~Arkani-Hamed, L.~Motl, A.~Nicolis and C.~Vafa,
	``The String landscape, black holes and gravity as the weakest force,''
	JHEP \textbf{06} (2007), 060
	doi:10.1088/1126-6708/2007/06/060
	[arXiv:hep-th/0601001 [hep-th]].
	
	\bibitem{Cheung:2018cwt}
	C.~Cheung, J.~Liu and G.~N.~Remmen,
	``Proof of the Weak Gravity Conjecture from Black Hole Entropy,''
	JHEP \textbf{10} (2018), 004
	doi:10.1007/JHEP10(2018)004
	[arXiv:1801.08546 [hep-th]].
	
	\bibitem{Heslop:2002hp}
	P.~J.~Heslop and P.~S.~Howe,
	``Four point functions in N=4 SYM,''
	JHEP \textbf{01} (2003), 043
	doi:10.1088/1126-6708/2003/01/043
	[arXiv:hep-th/0211252 [hep-th]].
	



\bibitem{fh}
F.~Aprile and P.~Heslop,
``Superconformal blocks in diverse dimensions and $BC$ symmetric functions,''
To appear.


\bibitem{website}
Per Alexandersson,
``Symmetricfunctions.com website,''
\href{https://www.symmetricfunctions.com/superSymmetricSchur.htm}{https://www.symmetricfunctions.com/superSymmetricSchur.htm}


\bibitem{shimeno}
N,~Shimeno,
``A formula for the hypergeometric function of type BCn,''
Pacific Journal of Mathematics,
\textbf{236(1)},105--118,(2008)
[arXiv:0706.3555]


\bibitem{grassmann}
M.~Headrick, \url{http://people.brandeis.edu/~headrick/Mathematica/grassmann.m}

\end{thebibliography}
\end{document}